\documentclass[10pt,twocolumn,letterpaper]{article}

\usepackage{wacv}              
\usepackage{times}  
\usepackage{helvet}  
\usepackage{courier}  
\usepackage[hyphens]{url}  
\usepackage{graphicx} 
\usepackage{natbib}  
\usepackage{caption} 
\usepackage[accsupp]{axessibility}  
\usepackage{epsfig}
\usepackage{graphicx}
\usepackage{amsmath}
\usepackage{amssymb}
\usepackage{enumitem}
\usepackage{tabularx}
\usepackage{booktabs}
\usepackage[linesnumbered,ruled]{algorithm2e}
\usepackage{algorithmic}
\usepackage{amsfonts}       
\usepackage{nicefrac}       
\usepackage{microtype}      
\usepackage[table]{xcolor}
\usepackage{multirow}
\usepackage{comment}
\usepackage[switch]{lineno}
\usepackage{dsfont}
\usepackage{booktabs}
\usepackage{siunitx}
\usepackage{soul}
\usepackage{upgreek}

\usepackage{makecell}

\definecolor{wacvblue}{rgb}{0.21,0.49,0.74}
\usepackage[pagebackref,breaklinks,colorlinks,allcolors=wacvblue]{hyperref}

\def\wacvPaperID{1889} 
\def\confName{WACV}
\def\confYear{2026}

\title{UltraClean: A Simple Framework to Train Robust Neural Networks against Backdoor Attacks}

\author{Bingyin Zhao\\
Pixocial Technology\\
{\tt\small bingyin.zhao@pixocial.com}
\and
Yingjie Lao\\
Tufts University\\
{\tt\small yingjie.lao@tufts.edu}
}

\begin{document}
\maketitle
\begin{abstract}
Backdoor attacks are emerging threats to deep neural networks, which typically embed malicious behaviors into a victim model by injecting poisoned samples. Adversaries can activate the injected backdoor during inference by presenting the trigger on input images. Prior defensive methods have achieved remarkable success in countering dirty-label backdoor attacks where the labels of poisoned samples are often mislabeled. However, these approaches do not work for a recent new type of backdoor -- clean-label backdoor attacks that imperceptibly modify poisoned data and hold consistent labels. More complex and powerful algorithms are demanded to defend against such stealthy attacks. In this paper, we propose UltraClean, a general framework that simplifies the identification of poisoned samples and defends against both dirty-label and clean-label backdoor attacks. Given the fact that backdoor triggers introduce adversarial noise that intensifies in feed-forward propagation, UltraClean first generates two variants of training samples using off-the-shelf denoising functions. It then measures the susceptibility of training samples leveraging the error amplification effect in DNNs, which dilates the noise difference between the original image and denoised variants. Lastly, it filters out poisoned samples based on the susceptibility to thwart the backdoor implantation. Despite its simplicity, UltraClean achieves a superior detection rate across various datasets and significantly reduces the backdoor attack success rate while maintaining a decent model accuracy on clean data, outperforming existing defensive methods by a large margin. Code is available at \url{https://github.com/bxz9200/UltraClean}.
\end{abstract}
    
\section{Introduction}
\label{sec:intro}

\begin{figure}[htbp]
    \centering
    \resizebox{1\linewidth}{!}{
    \includegraphics{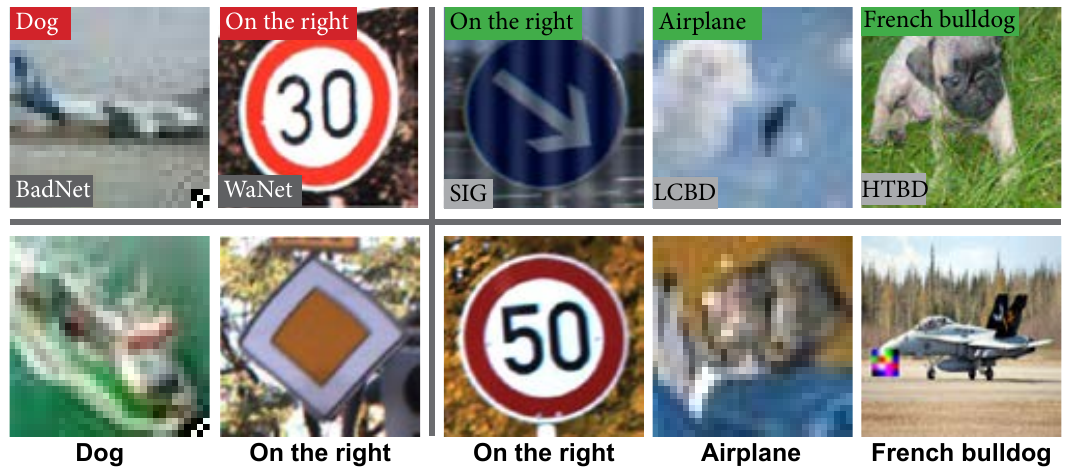}}
    \caption{Illustration of dirty-label and clean-label attacks (Top row: poisoned training samples; Bottom row: backdoored test samples. Red: incorrect labels; Green: correct labels. Dirty-label poisoned samples always possess incorrect labels while clean-label poisoned samples are imperceptible compared to benign samples and possess correct labels.} 
    \label{fig:dirty_clean_label_comparison}
    \vspace{-2em}
\end{figure}

With the thriving of machine learning, deep neural networks (DNNs) have achieved unprecedented progress and reached human-level performance in various tasks, including computer vision~\cite{he2016deep,DBLP:conf/iclr/DosovitskiyB0WZ21}, natural language processing~\cite{DBLP:conf/naacl/DevlinCLT19,DBLP:conf/nips/BrownMRSKDNSSAA20} and game playing~\cite{DBLP:journals/nature/SilverHMGSDSAPL16}. Given the remarkable success, DNNs are further deployed to safety-critical applications such as authentication~\cite{DBLP:journals/ijon/WangD21a} and autonomous driving~\cite{DBLP:journals/corr/BojarskiTDFFGJM16}. However, DNNs are proven to be vulnerable to a variety type of adversarial attacks. One notorious attack is the adversarial example~\cite{DBLP:journals/corr/GoodfellowSS14} that occurs at test-time, where an adversary fools well-trained DNN models by adding imperceptible perturbations on test images. Another well-known attack is the data poisoning attack~\cite{DBLP:conf/icml/BiggioNL12} that occurs in the training phase, where an adversary can inject well-crafted poisoned samples into the training data and introduce malicious behavior to models trained on the poisoned dataset. This paper focuses on backdoor attacks~\cite{gu2017badnets}, where an adversary attempts to contaminate the training dataset via data poisoning and install a backdoor into models trained on the corrupted dataset. During inference, backdoored models 

misclassify inputs with backdoor triggers to a target label while behaving normally on benign inputs.

Backdoor attacks come in two flavors: dirty-label attacks~\cite{DBLP:journals/corr/abs-1712-05526,DBLP:conf/cvpr/Moosavi-Dezfooli17,DBLP:conf/iclr/NguyenT21} and clean-label attacks~\cite{DBLP:conf/icip/BarniKT19,DBLP:journals/corr/abs-1912-02771,DBLP:conf/aaai/SahaSP20,DBLP:conf/cvpr/HanZCL0L25}. Poisoned samples of dirty-label attacks are always mislabeled. For example, as shown in Figure~\ref{fig:dirty_clean_label_comparison}, a poisoned sample ``airplane'' is labeled as ``dog''. In contrast, clean-label attacks hold consistent labels to images content. Although dirty-label attacks are effective, they can be easily distinguished due to incorrect labels. Existing defenses have shown decent performance in detecting and mitigating such attacks~\cite{DBLP:journals/corr/abs-2012-10544,DBLP:journals/corr/abs-2007-08745}. On the other hand, clean-label attacks are more stealthy and insidious. 
The clean-label poisoned samples are almost visually indistinguishable from benign samples; thus, defenses against the attack become a more challenging task. Some recent works~\cite{DBLP:journals/corr/abs-2110-11571,DBLP:journals/corr/abs-2202-03423} attempted to alleviate the clean-label attacks by decoupling the training phase to suppress the backdoor injection. However, they require complicated algorithms with significantly more operations in training and do not identify the backdoor samples in the poisoned dataset. Thus, users have to re-run the defense algorithms every time they train a new model by using the same potentially poisoned dataset, which dramatically increases the cost.

In this work, we propose UltraClean, a poisons-filtering-based framework (i.e., dataset cleanse) to detect backdoor samples, cleanse poisoned datasets, and train backdoor-mitigated models against both dirty-label and clean-label backdoor attacks. We focus on the image classification task. Our idea is inspired by prior works~\cite{DBLP:conf/iclr/LinGH19} and~\cite{DBLP:conf/cvpr/XieWMYH19,DBLP:conf/ndss/Xu0Q18}, which have demonstrated that adversarial perturbations of adversarial examples are amplified during the feed-forward propagation (i.e., \textbf{error amplification effect}) in DNNs and can be effectively eliminated by simple image-denoising techniques. We argue that although backdoor samples hold fundamentally different generation mechanisms to adversarial examples, the backdoor triggers share a similar error amplification effect as the adversarial perturbations. To this end, our training framework, UltraClean, employs off-the-shelf image-denoising functions and the error implication effect to filter out poisoned samples from benign training data and thwart the backdoor implantation.

The proposed method first trains an arbitrary model on the potentially poisoned dataset and uses it as the backdoor detection model. We then produce two variants of each training image using denoising functions and feed the difference corresponding to the original image into the detection model. We leverage the error amplification effect to compute the susceptibility of the training data in a feed-forward pass. The susceptibility of poisoned data tends to be higher, guiding the model to detect and remove these samples. Finally, we obtain a backdoor-mitigated model by retraining on the sterilized dataset. We show that UltraClean is highly effective in defending against various representative attacks including both dirty-label and clean-label ones, achieving a high detection rate while maintaining the model accuracy.

Our contributions are summarized as follows:
{\begin{itemize}[itemsep=0pt]
    \item UltraClean is a once-for-all backdoor-free training framework and a poisons-filtering-based (i.e., dataset cleanse) defense against both dirty and clean-label backdoor attacks. The method not only defends against backdoor attacks but also identifies poisoned samples regardless of generation mechanisms and dataset complexity.
    \item We propose an effective approach that does not affect the normal training process and significantly simplifies the backdoor identification and mitigation.
    \item We demonstrate that two widely adopted poisons-filtering-based defenses are only effective on dirty-label attacks but are unsuccessful in defending against clean-label attacks.
    \item We conduct comprehensive experiments and examine different types of dirty-label and clean-label attacks to illustrate the effectiveness of UltraClean. 
\end{itemize}}

\section{Related Work}
\label{sec:background}
\subsection{Backdoor Attacks on DNN}
Backdoor attacks can be broadly categorized into \textbf{data poisoning}~\cite{gu2017badnets,DBLP:journals/corr/abs-1712-05526,DBLP:conf/aaai/SahaSP20,zhao2022towards,zhao2022clpa} and \textbf{model poisoning}~\cite{DBLP:conf/nips/NguyenT20,DBLP:conf/iclr/NguyenT21,DBLP:conf/iccv/DoanL0L21,DBLP:conf/nips/DoanLL21}. Data poisoning backdoor attacks implant the backdoors by modifying training data (i.e., injecting poisoned samples), which does not require access to the training process. Such attacks affect victims indirectly through the poisoned data set. In contrast, model poisoning backdoor attacks require full control of the training process to alter model parameters and embed backdoors into the model. Such attacks affect victims by providing the poisoned model directly. \textbf{Our work focuses on the defense against data poisoning attacks via dataset cleanse.}

The very first backdoor attack via data poisoning against neural networks is BadNets~\cite{gu2017badnets} where poisoned training samples are generated by stamping a pre-defined pattern (trigger) onto benign images and assigned with incorrect target labels (dirty-label). One drawback of BadNets is that the poisoned samples are suspicious and quite easy to detect upon human inspection. Later on, a series of works are proposed to enhance the stealthiness by improving the invisibility of backdoor triggers. \cite{DBLP:journals/corr/abs-1712-05526} blends backdoor triggers with benign images; \cite{DBLP:conf/codaspy/ZhongLSZ020} imposes imperceptible perturbations onto poisoned samples using universal adversarial perturbations~\cite{DBLP:conf/cvpr/Moosavi-Dezfooli17}; \cite{DBLP:journals/tdsc/LiXZZZ21,li2021invisible} creates invisible triggers via steganography and regularization; \cite{DBLP:conf/aaai/0005LMZ21} employs controlled detoxification to perform the attack in the feature space. However, all these methods are still dirty-label attacks and could be detected by inspecting data labels.

On the other hand, clean-label backdoor attacks are more stealthy and hard to detect since they retain consistent labels to the image contents. For instance, \cite{DBLP:conf/icip/BarniKT19,DBLP:conf/eccv/LiuM0020,DBLP:conf/sp/QuiringR20} superimpose the trigger with benign images and employ different techniques (i.e., sinusoidal signal, reflection and image-scaling) to conceal the trigger; \cite{DBLP:journals/corr/abs-1912-02771} creates poisoned sample by placing a stealthy patch on hard-to-classify images generated by generative models or adversarial perturbations; \cite{DBLP:conf/aaai/SahaSP20} produces poisoned samples by minimizing their distance from source benign images in feature space. {The objective of UltraClean is to defend against both dirty-label attacks and clean-label attacks in a simple and effective fashion.} 



\subsection{Defenses}
Defenses against backdoor attacks are extensively studied in recent years, which can be broadly categorized as training-phase defense and inference-phase defense. Training-phase defense attempts to eliminate backdoor before model deployment by filtering poisoned training samples~\cite{DBLP:conf/aaai/ChenCBLELMS19,DBLP:conf/nips/Tran0M18,DBLP:conf/sp/ChouTP20,DBLP:conf/uss/Tang0TZ21,DBLP:conf/acsac/GaoXW0RN19,DBLP:conf/iccv/ZengPMJ21,DBLP:journals/corr/abs-2104-11315}, suppressing the effectiveness of poisoned data~\cite{DBLP:conf/iclr/DuJS20,DBLP:journals/corr/abs-2002-11497,DBLP:journals/corr/abs-2110-11571,DBLP:journals/corr/abs-2202-03423}, or reconstructing the trained model~\cite{zhao2018resilience,DBLP:conf/raid/0017DG18,DBLP:conf/iclr/ZhaoCDRL20,DBLP:conf/iclr/LiLKLLM21}. Inference-phase defense alleviates the backdoor effect after model deployment by pre-processing data~\cite{DBLP:conf/acsac/DoanAR20,udeshi2019model,qiu2021deepsweep}, filtering suspicious samples~\cite{javaheripi2020cleann,jin2020unified}, synthesizing possible triggers~\cite{wang2019neural,chen2019deepinspect,DBLP:conf/nips/QiaoYL19,DBLP:conf/icdm/GuoWXXDS20,DBLP:conf/mm/ZhuNWXW20}, or identifying if a model is backdoored~\cite{DBLP:journals/corr/abs-1911-07399,DBLP:conf/sp/XuWLBGL21,DBLP:conf/eccv/HuangPJT20,DBLP:conf/eccv/WangZLCXW20,DBLP:conf/cvpr/KolouriSPH20}. 

This paper proposes the first poisons-filtering-based defensive solution against a wide range of backdoor attacks (i.e., both dirty-label and clean-label). Our work is a training-phase defense where users seek to detect and remove malicious training samples and train a backdoor-free model. The most relevant works are the spectral signatures defense~\cite{DBLP:conf/nips/Tran0M18} and strong intentional perturbation defense~\cite{DBLP:conf/acsac/GaoXW0RN19} that can effectively distinguish poisoned samples in the dataset. However, we find that these defenses are only effective against dirty-label attacks. 

\section{UltraClean}
\label{sec:methods}


\subsection{Threat Model}
We consider the same threat model defined in prior data poisoning backdoor works~\cite{DBLP:conf/nips/Tran0M18} where the adversary targets to implant a backdoor into a DNN for an image classification task by injecting a fraction of poisoned samples into the training dataset. We assume a strong adversary that knows standard training algorithms, classic model architectures, and the statistical information of the training dataset to craft powerful and stealthy poisoned samples. However, the model is not trained by the adversary but the user with possibly contaminated training dataset from an uncertified source. Under the threat model, we evaluate the effectiveness of the proposed defensive framework from three dimensions.

\textbf{Backdoor detection rate (BDR).} The detection rate is the fraction of poisoned samples detected by the defense. We seek to detect as many poisoned samples as possible.

\textbf{Attack success rate (ASR).} The ASR is the fraction of test images classified as the target label in the presence of the backdoor trigger. We want the ASR to be as low as possible after retraining on the sanitized dataset.

\textbf{Model accuracy on clean data.} The model accuracy is the fraction of benign images that the trained model correctly classifies. We hope the accuracies before and after retraining remain as close as possible.



\begin{algorithm}[htbp]
\DontPrintSemicolon 
\KwIn{Training dataset $(\mathbf{x},\mathbf{y})$ $\sim$ $\boldsymbol{D_{tr}}$, \\ \quad \quad \quad Randomly initialized deep neural network model $\mathbf{M_{\theta}}$, \\ \quad \quad \quad Detection and removal threshold $\boldsymbol{{\beta}}$,\\  \quad \quad \quad Output of a potentially backdoored model $\boldsymbol{f_{\theta^{*}}(\cdot)}$, \\ \quad \quad \quad Total number of training samples $\boldsymbol{n}$},
\KwOut{Sanitized training dataset $\boldsymbol{D_{s}}$, \\ \quad \quad \quad \quad  Post-clean model $\mathbf{M_{\hat{\theta}}}$}

\textit{\#Train the model on the potentially poisoned dataset}

\textbf{Minimize} $\mathcal{L} = \mathds{E}_{(\mathbf{x},\mathbf{y})\sim \boldsymbol{D_{tr}}}[\ell(\mathbf{M_{\theta}}(\mathbf{x}),\mathbf{y})]\rightarrow \mathbf{M_{\theta^{*}}}$ 

\textit{\#Initialize the sanitized training set and score list}

$\boldsymbol{D_{s}}$ $\leftarrow$ \{\}; \\ $S$ $\leftarrow$ MaxHeap []


\textit{\#Generate denoised variants}

\textbf{for all $\mathbf{x}$ do} 

\quad \textbf{for $c_i$ in $\mathbf{x}$ do} 

\quad\quad $\tilde{{c_i}}$ =$ \frac{1}{C(p)}\sum_{\substack{q\in \Omega(p,r)}} c_i(q)w(p,q) \rightarrow \tilde{\mathbf{x_1}}$

\quad\quad $\tilde{c_i} = median\{q\in \Omega(p):c_i(q)\}\rightarrow \tilde{\mathbf{x_2}}$

\quad \textbf{end for}

\quad \textit{\#Enlarge error using error amplication effect} 

\quad $\mathbf{v} = \boldsymbol{f_{\theta^{*}}(\mathbf{x})}$; $\mathbf{v_1} = \boldsymbol{f_{\theta^{*}}(\tilde{\mathbf{x_1}})}$; $\mathbf{v_2} = \boldsymbol{f_{\theta^{*}}(\tilde{\mathbf{x_2}})}$

\quad \textit{\#Compute susceptibility} 

\quad $\mathbf{s} = ||\mathbf{v}-\mathbf{v_1}||_1 + ||\mathbf{v}-\mathbf{v_2}||_1$ 


\quad \textbf{heappush}$(S, (\mathbf{s}, (\mathbf{x},\mathbf{y})))$

\textbf{end for}

\textit{\#Remove poisoned samples from the dataset} 

\textbf{for} i = 0 \textbf{to} $\boldsymbol{{\beta}}\cdot\boldsymbol{n}$ \textbf{do} 

\quad  $\boldsymbol{D_{tr}}$.remove(\textbf{heappop}(S)[1]) $\rightarrow \boldsymbol{D_s}$

\textbf{end for}


\textit{\#Train/Retrain the model on the sterilized dataset}

\textbf{Minimize} $\mathcal{L} = \mathds{E}_{(\mathbf{x},\mathbf{y})\sim \boldsymbol{D_{s}}}[\ell(\mathbf{M_{\theta}}(\mathbf{x}),\mathbf{y})]\rightarrow \mathbf{M_{\hat{\theta}}}$ 

Return $\mathbf{M_{\hat{\theta}}}, \boldsymbol{D_s}$.
\caption{UltraClean}
\label{Algorithm}
\end{algorithm}

\begin{figure*}[htbp]
    \centering
    \resizebox{0.8\textwidth}{!}{
    \includegraphics{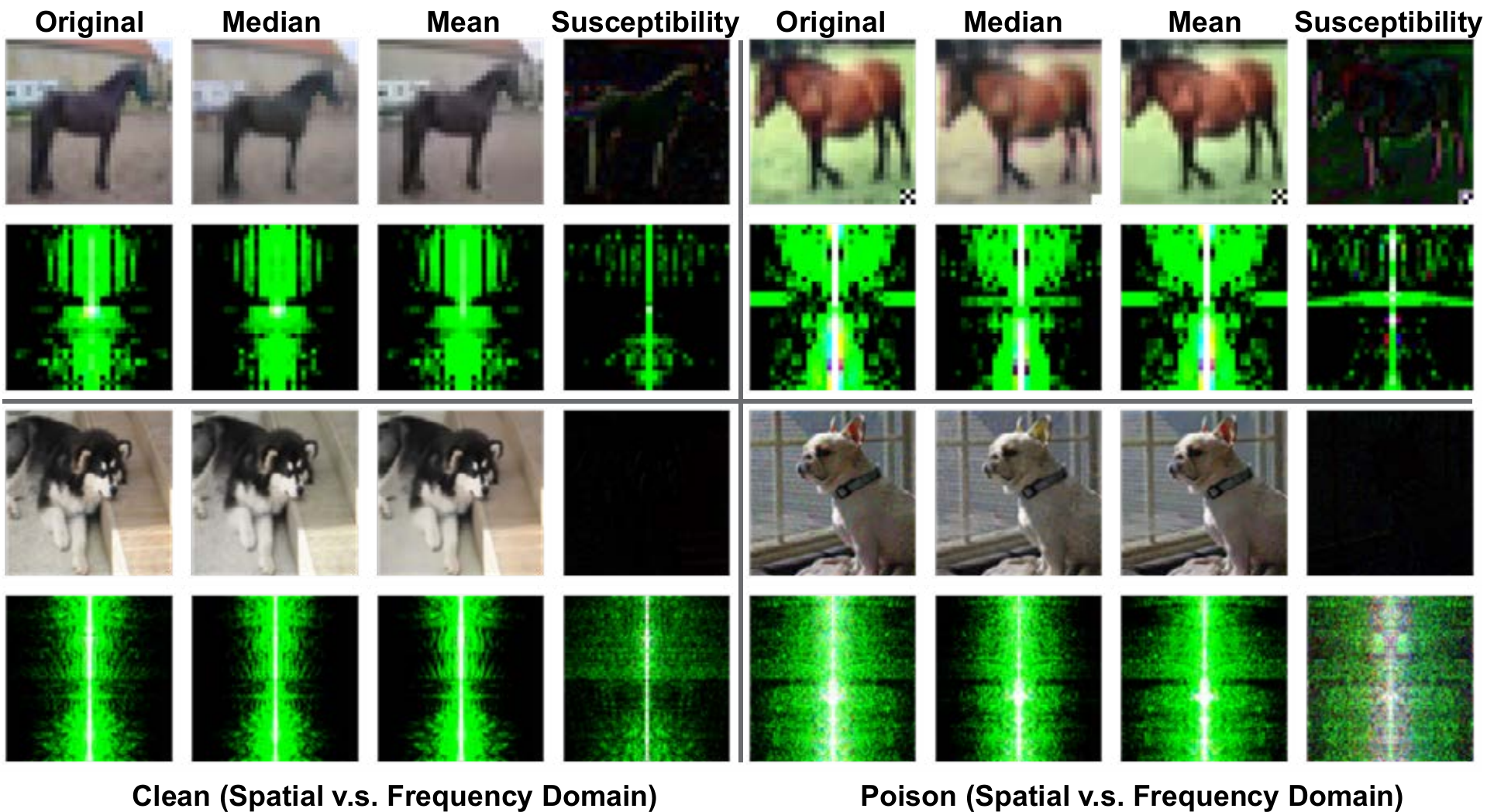}}
    \caption{Spatial and frequency domain views of two backdoor attacks. Top two rows are the dirty-label attack (BadNets~\protect\cite{gu2017badnets}); Bottom two rows are the clean-label attack (Hidden Trigger Backdoor~\protect\cite{DBLP:conf/aaai/SahaSP20}). As shown in the 4th and 8th columns, poisons reveal substantial qualitative noise difference than the clean counterparts. The noise difference in pixel space are amplified during the feed-forward propagation in a deep neural network and become a strong indicator to differentiate poisoned and benign samples.}
    \label{fig:Explanation_UltraClean}
\end{figure*}

\subsection{UltraClean Framework}

The proposed UltraClean is a general framework that aims at defending against various backdoor attacks by profoundly cleansing poisoned samples from the training dataset. UltraClean consists of three phases: pre-clean training, poisons clean (detection and removal), and post-clean retraining. In the pre-clean training, we train the DNN model on the possibly poisoned dataset. Note that the trained model will be mounted with the backdoor as intended. In the poisons clean phase, we first generate two baseline images of the training data using two denoising filters separately and then compute the $\ell_1$-norm distance score of softmax layer output between the original image and its pair of denoised variants using the DNN model obtained in the pre-clean training phase. The denoised-original distance scores (susceptibility) of poisoned samples tend to be higher than benign samples because the error is amplified during feed-forward propagation. Thus, the poisoned samples can be easily detected and removed from the training dataset upon the susceptibility. Lastly, we retrain the model on the sanitized training dataset and acquire the backdoor-mitigated model. 
The concrete account of the UltraClean framework is presented in Algorithm~\ref{Algorithm}. 


\subsection{Methodology}
The key of UltraClean is to develop an approach that can differentiate poisoned and benign samples, which hence can be effective under both dirty-label and clean-label settings. In general, backdoor attacks require a trigger to induce DNNs to learn the malicious behavior. Such a trigger, whether visually perceptible or not, introduces extra noise (pixels that do not match the image content) to benign images. We plot the frequency-domain view in Figure~\ref{fig:Explanation_UltraClean} to visualize the signal distribution in images. The central area of the frequency-domain figure represents the low-frequency signal, while the peripheral area represents the high-frequency signal. As shown in the 1st and 5th columns, poisons demonstrate more high-frequency noise than benign data. The noise is bounded by the maximum perturbation added to the images and is sometimes hard to detect in the pixel space (i.e., some works attempt to minimize the perturbation~\cite{DBLP:journals/corr/abs-1912-02771,DBLP:conf/aaai/SahaSP20}). For example, it is barely impossible to perceptually distinguish the poison sample (the right side of the 3rd row) from the clean sample (the left side of the 3rd row) while they have obvious differences in the frequency domain (the 4th and 8th columns in the 4th row). 
Moreover, the noise distributions vary due to different mechanisms of poisoned sample generation, which demands high generalizability of the detection.

To tackle this issue, we propose a simple yet effective approach that differentiates poisoned samples by measuring the susceptibility leveraging the \textbf{error amplification effect} of DNN along with two off-the-shelf denoising functions. We choose non-local mean and local median as the denoising functions based on the following reasons: i). We want to keep our design simple and effective so that UltraClean can be applied in more general cases; ii). The non-local mean and the local median are the most common denoising techniques and can be implemented without additional cost to the training process; iii). The poisons-filtering algorithm should be agnostic to the mechanism of poisoned sample generation. While other denoising techniques might also be used in our framework, these two denoising methods complement each other (i.e., the mean filter is linear and the median filter is non-linear) and are surprisingly effective in empirically identifying backdoor samples against various backdoor triggers when incorporated with the error amplification effect. We briefly introduce the denoising functions below and conduct ablation studies to illustrate their importance and complementarity in backdoor detection in the supplementary materials.

\textbf{Non-local mean denoising}~\cite{DBLP:conf/cvpr/BuadesCM05} removes noise by replacing each pixel value with a weighted mean computed over global spatial regions. It is defined as:
\begin{equation}
    \tilde{c}(p) = \frac{1}{C(p)}\int g(d(\Omega(p,r),\Omega(q,r)))c(q)dq,
\label{equ:non-local-mean-V1}
\end{equation}
where $\tilde{c}(p)$ is the denoised value of pixel $p$, $d(\cdot)$ represents the Euclidean distance between spatial regions $\Omega(p,r)$ and $\Omega(q,r)$. $\Omega(p,r)$ and $\Omega(q,r)$ are search windows centered at pixels $p$ and $q$, and the boundary of which is defined by $r$. $C(\cdot)$ is a normalization function and $g(\cdot)$ is a decreasing function.
In this work, we exploit the pixel-wise implementation of the algorithm. Consider a color image with RGB channels as $\mathbf{x} = (c_1, c_2, c_3)$, the denoising operation is expressed as:
\begin{equation}
\label{equ:non-local-mean-V2}
\begin{aligned}
    \tilde{c_i}(p) &= \frac{1}{C(p)}\sum_{\substack{q\in \Omega(p,r)}} c_i(q)w(p,q), \\
    &C(p) = \sum_{\substack{q\in \Omega(p,r)}} w(p,q),
\end{aligned}
\end{equation}
where $w(p,q)$ is a weighting function depends on the squared distance $d^{2}(\cdot)$ as expressed in Equation~(\ref{equ:distance-function}), which can be calculated by 
Equation~(\ref{equ:kernel}).
\begin{equation}
    d^{2} = \frac{\sum_{i=1}^{3}\sum_{j\in\Omega(0,r)}(c_i(p+j)-c_i(q+j))^2}{3\times(2r+1)^2}.
\label{equ:distance-function}
\end{equation}
\begin{equation}
    w(p,q) = e ^{-\frac{max(d^2-2\sigma^2,0.0)}{h^2}}.
\label{equ:kernel}
\end{equation}

\textbf{Local median denoising} removes noise by weighting nearby pixels of each pixel using the median smoothing filter. The filter scans over each pixel $p$ and replaces the value of the center pixel with the median value of surrounding pixels, which can be expressed as:
\begin{equation}
    \tilde{c_i}(p) = median\{q\in \Omega(p):c_i(q)\}.
\label{equ:local-meidan}
\end{equation}

We acknowledge there exist many other denoising algorithms designed specifically for certain types of noise and conduct an abliation study in Appendix~\ref{sec: ablation_denoising}, where we observe that using the aforementioned two denoising techniques can already achieve excellent performance with trivial complexity.  
The workflow of UltraClean is straightforward, for each training image $\mathbf{x}$, we generate two denoised versions $\tilde{\mathbf{x_1}}$ (e.g., the 2nd and 6th columns in Fig~\ref{fig:Explanation_UltraClean}) and $\tilde{\mathbf{x_2}}$ (e.g., the 3rd and 7th columns). $\tilde{\mathbf{x_1}}$ and $\tilde{\mathbf{x_2}}$ serve as baselines to compute the susceptibility. We define susceptibility as the error (i.e., noise difference) between original images and denoised variants. It can be seen from the 4th and 8th columns that the error of poisoned images are considerably higher than that of clean images. However, as seen in the 6th and 7th columns, the denoising functions do not significantly reduce the noise in the denoised variants compared to the original poisoned samples (the 5th column), so we are not able to distinguish poisoned samples from benign samples by simply computing the susceptibility in the pixel space. 

We propose a method to enlarge the error to facilitate the computation using the error amplification effect, which is a property of DNNs that minor adversarial perturbations in model inputs accumulate during forward propagation and affect the model outputs~\cite{DBLP:conf/iclr/LinGH19}. 
We feed the original image and denoised variants into the pre-trained DNN and obtain three vectors $\mathbf{v}$, $\mathbf{v_1}$, and $\mathbf{v_2}$, which are the softmax outputs of the model. We compute the $\ell_1$-norm distances of ($\mathbf{v}$, $\mathbf{v_1}$) and ($\mathbf{v}$, $\mathbf{v_2}$), respectively, and then aggregate the results to obtain the amplified noise difference. The susceptibility can thus be mathematically defined as:
\begin{equation}
    \mathbf{s}= ||\mathbf{v}-\mathbf{v_1}||_1 + ||\mathbf{v}-\mathbf{v_2}||_1
    \label{equ:susceptibility}
\end{equation}
Finally, we obtain the sanitized dataset by removing a portion of samples with the highest susceptibility. Note that users just need to run UltraClean once and any model trained on the cleansed dataset will be backdoor-free.

\section{Experiments}
\label{sec:experiments}


\begin{table*}[htbp]
\centering
\small
\renewcommand\arraystretch{0.5}
\begin{tabular}{
m{1.1cm}<{\centering}
|m{1.0cm}<{\centering}
m{1.0cm}<{\centering}
m{1.0cm}<{\centering}
m{1.0cm}<{\centering}
m{1.0cm}<{\centering}
m{1.0cm}<{\centering}
m{1.0cm}<{\centering}
m{1.0cm}<{\centering}
}
\hline 
\makecell[c]{\textbf{Class}\\ \textbf{(ID)}} &
\makecell[c]{\textbf{Acc.}\\ \textbf{(PC)}} &
\makecell[c]{\textbf{ASR}\\ \textbf{(PC)}} &
\makecell[c]{\textbf{Acc.}\\ \textbf{(SVD)}} &
\makecell[c]{\textbf{ASR}\\ \textbf{(SVD)}} &
\makecell[c]{\textbf{BDR}\\\textbf{(SVD)}} &
\makecell[c]{\textbf{Acc.}\\\textbf{(UC)}} &
\makecell[c]{\textbf{ASR}\\\textbf{(UC)}} &
\makecell[c]{\textbf{BDR}\\\textbf{(UC)}} \\

\hline       
\includegraphics[scale=0.05]{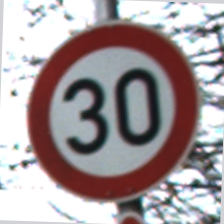}(0) &95.83\% &47.58\%  &96.18\%  &58.36\% &43.48\% & {96.52\%} &\textbf{8.18\%} &\textbf{52.17\%} \\ 
\includegraphics[scale=0.05]{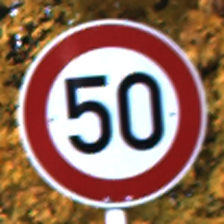}(1) &97.22\% &54.15\% &{96.18\%} &62.82\% &44.87\%  &95.49\% &\textbf{10.47\%} &\textbf{56.41\%} \\
\includegraphics[scale=0.05]{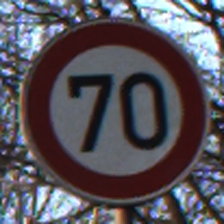}(2) &95.83\% &67.62\% &96.52\%  &\textbf{26.33\%} &\textbf{67.21\%}   &{96.87\%} &61.92\% &50.82\%\\ 
\includegraphics[scale=0.05]{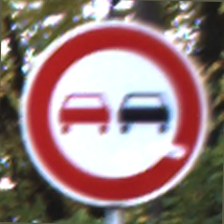}(3) &95.14\% &17.38\% &{97.22\%}  &2.13\% &50.99\%  &95.14\% &\textbf{0.00\%} &\textbf{69.09\%}  \\ 
\includegraphics[scale=0.05]{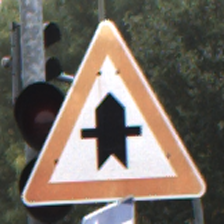}(4) &95.49\% &51.53\% &95.49\%  &{40.46\%} &\textbf{57.43\%}  &{96.18\%} &\textbf{1.91\%} &46.62\%\\ 
\includegraphics[scale=0.05]{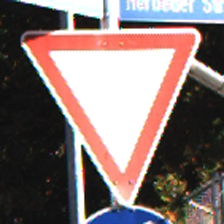}(5) &97.57\% &51.78\% &{95.83\%}  &53.75\% &42.56\%  &93.05\% &\textbf{37.15\%} &\textbf{52.82\%} \\ 
\includegraphics[scale=0.05]{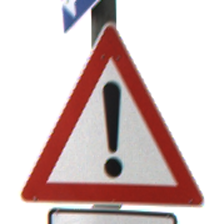}(6) &95.49\% &74.16\% &96.18\%  &62.92\% &40.16\% &{96.52\%} &\textbf{17.98\%} &\textbf{62.99\%} \\   
\includegraphics[scale=0.05]{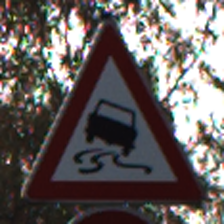}(7) &95.83\% &63.83\% &96.88\%  &62.92\% &30.77\%   &{97.22\%} &\textbf{55.32\%} &\textbf{51.92\%} \\ 
\includegraphics[scale=0.05]{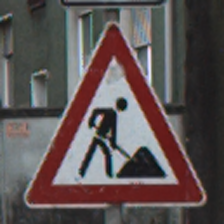}(8) &97.92 &73.41\% &{96.18\%}  &75.79\% &40.70\%   &{96.18\%} &\textbf{40.08\%} &\textbf{48.74\%}\\   
\includegraphics[scale=0.05]{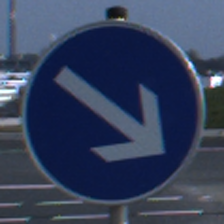}(9) &97.57\% &47.18\% &93.06\%  &22.54\% &64.58\%   &{97.22\%} &\textbf{6.34\%} &\textbf{91.66\%}\\ 

 \hline 
\end{tabular}
\caption{Comparison of accuracy ASR and BDR against SIG attack on GTSRB (SVD v.s. UltraClean). Numbers in the parenthesis are poisoned class IDs}
\vspace{-0.5em}
\label{detection_rate_sig}
\end{table*}

\subsection{Experiment Settings}
We consider multiple representative dirty-label and clean-label attacks within the threat model. Please note that attacks that disable training phase defense where attackers train the models do not align with our threat model. For dirty-label attacks, we implement BadNets~\cite{gu2017badnets}, Blended and Random Pattern~\cite{DBLP:journals/corr/abs-1712-05526}, and Trojan~\cite{Trojannn} following the original papers, and evaluate the effectiveness of UltraClean on CIFAR-10. For clean-label attacks, we implement Sinusoidal Signal Backdoor (SIG)~\cite{DBLP:conf/icip/BarniKT19}; Label Consistent Backdoor (LCBD)~\cite{DBLP:journals/corr/abs-1912-02771} and Hidden Trigger Backdoor (HTBD)~\cite{DBLP:conf/aaai/SahaSP20} based on their open-source repositories, and adopt the same training algorithms and dataset as in original papers for assessment (i.e., GTSRB for SIG, CIFAR-10 for LCBD, and ImageNet for HTBD). The mechanisms of attacks, detailed training settings, statistics of datasets, and neural network architectures are also summarized in the supplementary materials.

\vspace{-0.5em}

\subsection{Evaluation on Dirty-Label Attacks}
We first demonstrate UltraClean's effectiveness against various dirty-label attacks and refer to the SOTA defense against such attacks -- the Frequency Detection (FD)~\cite{DBLP:conf/iccv/ZengPMJ21} as the baseline for comparison. Note that FD is ineffective against clean-label attacks, as shown in~\cite{DBLP:conf/iccv/ZengPMJ21}. We follow the original poisoning practice and set the blended injection ratio to 0.2 and Trojan transparency to 0.5, respectively. The results are presented in Table~\ref{dirty-label-attacks}. UltraClean has an average of 97.78\% detection rate and an average of 97.93\% ASR reduction at a removal threshold of 0.3, achieving comparable performance to FD and indicating a strong capability of defending against dirty-label attacks. {In practice, users usually do not know how many poisoned samples are injected. The selection of $\beta$ should depend on the specific requirement of the application. Under safety-critical scenarios, we consider grid search a practical and systematic algorithm to determine the best value of $\beta$ given an accuracy threshold.}
\vspace{-0.5em}

\begin{table}[htbp]
\centering
\setlength{\tabcolsep}{1pt}
\resizebox{\linewidth}{!}{
\begin{tabular}{cccccccc}
\hline
\textbf{\begin{tabular}[c]{@{}c@{}}Attack\\ Type\end{tabular}} &
\textbf{\begin{tabular}[c]{@{}c@{}}Acc.\\ (PC)\end{tabular}} & \textbf{\begin{tabular}[c]{@{}c@{}}ASR\\ (PC)\end{tabular}} & \textbf{\begin{tabular}[c]{@{}c@{}}Acc.\\ (UC)\end{tabular}} & \textbf{\begin{tabular}[c]{@{}c@{}}ASR\\ (UC)\end{tabular}} &  \textbf{\begin{tabular}[c]{@{}c@{}}\ BDR\\ (UC)\end{tabular}} &  \textbf{\begin{tabular}[c]{@{}c@{}}\ BDR\\ (FD)\end{tabular}}  \\ \hline
\textbf{BadNets}                                                                                                           & 85.27\%                                                           & 100.00\%                                                                                                                                 & 83.91\%                                                               & \textbf{0.83\%}  & \textbf{94.42\%} & 90.50\%                                                            \\

\textbf{Trojan}                                                                                                                      &  84.94\%                                                                  & 99.19\%                                                                                                                                   &  84.73\%                                                              &   \textbf{1.61\%}       &  99.50\%  &  \textbf{99.99\% }                                                    \\
\textbf{Blended (HK)}                                                                                                                      &  84.99\%                                                                  & 97.47\%                                                                                                                                   &  85.08\%                                                              &   \textbf{3.06\%}      &  \textbf{97.38\%}  &  96.30\%                                                      \\
\textbf{Blended (RP)}                                                                                                       & 86.24\%                                                              & 99.96\%                                                                                                                                   &  84.23\%                                                              &   \textbf{0.15\%} &  \textbf{99.80\% }&  96.30\%                                                            \\ 

\hline

\end{tabular}
}
\caption{Performance comparison of UltraClean and prior SOTA Frequency Detection against dirty-label attacks.}
\label{dirty-label-attacks}
\vspace{-0.5em}
\end{table}

\subsection{Evaluation on Clean-Label Attacks}
We are more interested in the performance of UltraClean on clean-label attacks since there is no known effective poisons-filtering-based defense against such stealthy attacks. Thus, we comprehensively evaluate the performance of UltraClean against clean-label attacks. We adapt SVD~\cite{DBLP:conf/nips/Tran0M18} and STRIP~\cite{DBLP:conf/acsac/GaoXW0RN19} as the baseline methods for comparison. We follow the works of SVD and STRIP, and consider two different scenarios:
\begin{itemize}[itemsep=0pt,topsep=0pt]
    \item[1)] \textbf{detection on the poisoned class} where we assume that the defender already knows the poisoned class (i.e., target class);
    \item[2)] \textbf{detection on the whole training dataset} where the defender does not have knowledge of the data poisoning process. 
\end{itemize} 
We first present the results of detection on the poisoned class and then the detection on the whole training dataset. The later scenario is more practical and challenging.

\subsection{Detection on the Poisoned Class}

\textbf{SIG on GTSRB.} We follow the original recipe of SIG and set the frequency $f = 6$ and the strength $\Delta = 20$ to craft poisoned samples. To comprehensively evaluate the effectiveness of UltraClean, we iterate through each class as the target class and inject 30\% poisoned samples to the target class. For a fair comparison, we hold the same removal threshold as SVD. We present the post-clean accuracy, ASR and BDR of classes with the top 10 attack success rates in Table~\ref{detection_rate_sig}. Our proposed method achieves a higher detection rate in all classes except classes 2 and 4, outperforming SVD by a large margin in general. Meanwhile, UltraClean and SVD do not undermine the model performance after retraining on the sanitized dataset. 
Note that although $\Delta$ is fixed during training, the adversary can raise the signal strength to achieve better ASR at test time. Therefore, for the evaluation of ASR, we set $\Delta$ to 80 to achieve the best attack performance. It can be seen that UltraClean significantly reduces the post-clean ASR and outperforms SVD in all classes except class 2. For classes 3, 4, and 9, UltraClean can even achieve nearly 0\% post-clean ASR. 

\textbf{LCBD on CIFAR-10.} We pick class ``airplane'' as the target class and poison the dataset with 4\% of the entire images. LCBD uses two approaches to craft hard-to-classify images. For the GAN-based method, parameter $\uptau$ controls the interpolation between two images. For the AE-based method, parameter $\epsilon$ is the maximum perturbation added on images in $\ell_p$-norm. The adversary can construct various poisoned samples by varying these parameters. We present the settings that achieve the best attack performance and summarize the results of detection rate, ASR, and model accuracy in Table~\ref{LCBD_performance}. UltraClean shows superior performance on the defense against LCBD, achieving a much higher detection rate and lower ASR than SVD under all the settings. Meanwhile, UltraClean also retains decent post-clean model accuracy. An interesting phenomenon is that the post-clean ASR of SVD is even worse than the pre-clean ASR, which was also revealed in the SIG experiments. We argue the rationale behind this is that SVD removes less poisoned samples and more benign samples, rendering a higher percentage of poisoned samples in the entire dataset after detection. Another possible reason is that the poisoned samples removed by UltraClean may play a more critical role in embedding the backdoor than those removed by SVD. 

\begin{table}[htbp]
\setlength{\tabcolsep}{10pt}
\resizebox{\linewidth}{!}{
\begin{tabular}{c|ccc}
\hline
\textbf{Attack Type} & \textbf{Acc.} & \textbf{ASR}     & \textbf{BDR}     \\ \hline
\textbf{}            & \multicolumn{3}{c}{\textbf{Pre-clean (PC)}}         \\ \cline{2-4} 
\textbf{GAN ($\uptau$ = 0.3)}         & 88.17\%       & 83.03\%          & /                \\ 
\textbf{AE ($\ell_2$, $\epsilon$ = 1200)}          & 87.73\%       & 99.98\%          & /                \\ 
\textbf{AE ($\ell_\infty$, $\epsilon$ = 32)}          & 87.84\%       & 97.20\%          & /                \\ \hline
\textbf{}            & \multicolumn{3}{c}{\textbf{SVD}}                    \\ \cline{2-4} 
\textbf{GAN ($\uptau$ = 0.3)}         & 86.54\%       & 97.96\%          & 52.05\%          \\ 
\textbf{AE ($\ell_2$, $\epsilon$ = 1200)}          & 87.17\%       & 99.72\%          & 80.95\%          \\ 
\textbf{AE ($\ell_\infty$, $\epsilon$ = 32)}          & 86.97\%       & 99.82\%          & 69.85\%          \\ \hline
\textbf{}            & \multicolumn{3}{c}{\textbf{UltraClean (UC)}}        \\ \cline{2-4} 
\textbf{GAN ($\uptau$ = 0.3)}         & 86.98\%       & \textbf{26.94\%} & \textbf{79.75\%} \\
\textbf{AE ($\ell_2$, $\epsilon$ = 1200)}          & 87.26\%       & \textbf{1.10\%}  & \textbf{98.55\%} \\
\textbf{AE ($\ell_\infty$, $\epsilon$ = 32)}          & 87.60\%       & \textbf{1.13\%}  & \textbf{97.15\%} \\ \hline
\end{tabular}}
\caption{Comparison of backdoor detection and mitigation performance against LCBD attack on CIFAR-10.}
\label{LCBD_performance}
\end{table}

\textbf{HTBD on ImageNet.} For the HTBD attack, we follow the same settings in the original paper~\cite{DBLP:conf/aaai/SahaSP20} and conduct the experiment on 10 randomly selected target classes. A total of 100 poisoned samples are injected into the target class. Table~\ref{HTBD_performance} illustrates the detection rate, ASR and model accuracy. It can be seen that for most classes, poisoned samples completely bypass the SVD defense. However, UltraClean captures almost all poisoned samples and significantly undermines the backdoor effect to nearly 0\% ASR. Moreover, UltraClean even improves the post-clean model accuracy.

\begin{table}[htbp]
\setlength{\tabcolsep}{8pt}
\resizebox{\linewidth}{!}{
\begin{tabular}{cc|ccc}
\hline
\multicolumn{2}{c|}{\textbf{Target Class}}      & \textbf{\begin{tabular}[c]{@{}c@{}}Acc.\\ (PC)\end{tabular}} & \textbf{\begin{tabular}[c]{@{}c@{}}ASR\\ (PC)\end{tabular}} & \textbf{\begin{tabular}[c]{@{}c@{}}BDR\\ (SVD*)\end{tabular}} \\ \hline
\multicolumn{1}{c|}{\textbf{1}}  & Terrier      & 96.00\%                                                      & 45.25\%                                                     & 0.00\%                                                        \\
\multicolumn{1}{c|}{\textbf{2}}  & Bee          & 97.00\%                                                      & 75.00\%                                                     & 0.00\%                                                        \\
\multicolumn{1}{c|}{\textbf{3}}  & Plunger      & 95.00\%                                                      & 74.50\%                                                     & 55.00\%                                                       \\
\multicolumn{1}{c|}{\textbf{4}}  & Partridge    & 97.00\%                                                      & 87.75\%                                                     & 0.00\%                                                        \\
\multicolumn{1}{c|}{\textbf{5}}  & Ipod         & 95.00\%                                                      & 44.50\%                                                     & 8.00\%                                                        \\
\multicolumn{1}{c|}{\textbf{6}}  & Deerhound    & 95.00\%                                                      & 83.75\%                                                     & 0.00\%                                                        \\
\multicolumn{1}{c|}{\textbf{7}}  & Cockatoo     & 96.00\%                                                      & 78.00\%                                                     & 0.00\%                                                        \\
\multicolumn{1}{c|}{\textbf{8}}  & Toyshop      & 95.00\%                                                      & 80.50\%                                                     & 0.00\%                                                        \\
\multicolumn{1}{c|}{\textbf{9}}  & Tiger beetle & 98.00\%                                                      & 58.00\%                                                     & 0.00\%                                                        \\
\multicolumn{1}{c|}{\textbf{10}} & Goblet       & 95.00\%                                                      & 89.75\%                                                     & 0.00\%                                                        \\ \hline
\multicolumn{2}{c|}{\textbf{Target Class}}      & \textbf{\begin{tabular}[c]{@{}c@{}}Acc.\\ (UC)\end{tabular}} & \textbf{\begin{tabular}[c]{@{}c@{}}ASR\\ (UC)\end{tabular}} & \textbf{\begin{tabular}[c]{@{}c@{}}BDR\\ (UC)\end{tabular}}   \\ \hline
\multicolumn{1}{c|}{\textbf{1}}  & Terrier      & 100.00\%                                                     & \textbf{0.00\%}                                             & \textbf{97.00\%}                                              \\
\multicolumn{1}{c|}{\textbf{2}}  & Bee          & 99.00\%                                                      & \textbf{0.00\%}                                             & \textbf{86.00\%}                                              \\
\multicolumn{1}{c|}{\textbf{3}}  & Plunger      & 97.00\%                                                      & \textbf{4.75\%}                                             & \textbf{86.00\%}                                              \\
\multicolumn{1}{c|}{\textbf{4}}  & Partridge    & 100.00\%                                                     & \textbf{0.00\%}                                             & \textbf{87.00\%}                                              \\
\multicolumn{1}{c|}{\textbf{5}}  & Ipod         & 100.00\%                                                     & \textbf{0.00\%}                                             & \textbf{96.00\%}                                              \\
\multicolumn{1}{c|}{\textbf{6}}  & Deerhound    & 100.00\%                                                     & \textbf{0.25\%}                                             & \textbf{96.00\%}                                              \\
\multicolumn{1}{c|}{\textbf{7}}  & Cockatoo     & 98.00\%                                                      & \textbf{0.00\%}                                             & \textbf{99.00\%}                                              \\
\multicolumn{1}{c|}{\textbf{8}}  & Toyshop      & 98.00\%                                                      & \textbf{4.50\%}                                             & \textbf{72.00\%}                                              \\
\multicolumn{1}{c|}{\textbf{9}}  & Tiger beetle & 100.00\%                                                     & \textbf{0.00\%}                                             & \textbf{100.00\%}                                             \\
\multicolumn{1}{c|}{\textbf{10}} & Goblet       & 99.00\%                                                      & \textbf{27.50\%}                                            & \textbf{86.00\%}                                              \\ \hline
\end{tabular}}
\caption{Comparison of clean accuracy, ASR and BDR against HTBD attack on ImageNet (``*'' denotes results replicated from~\cite{DBLP:conf/aaai/SahaSP20}).}
\label{HTBD_performance}
\vspace{-1em}
\end{table}

\subsection{Detection on the Whole Training Dataset}

\begin{figure*}[htbp]
    \centering
    \resizebox{0.9\textwidth}{!}{
    \includegraphics{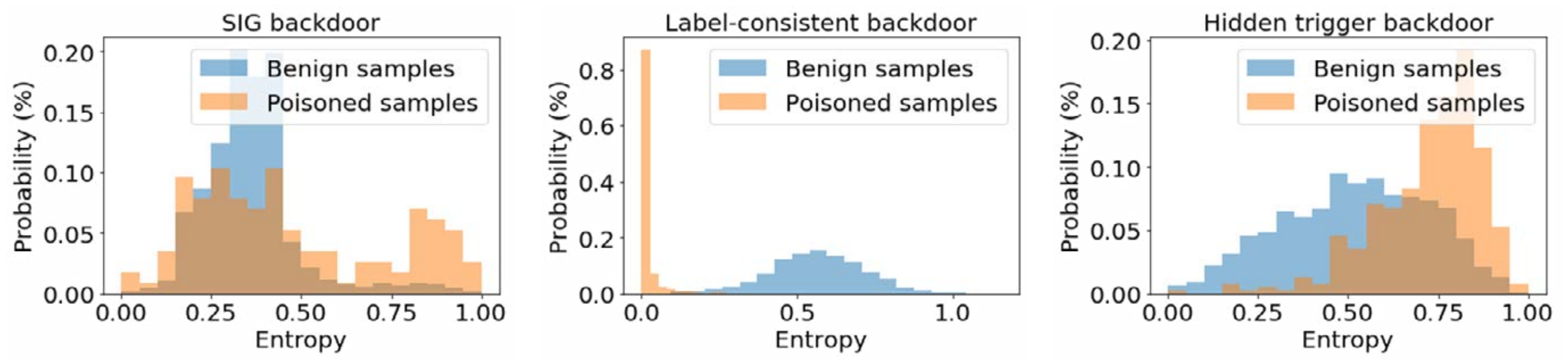}}
    \caption{STRIP against SIG backdoor (left), LCBD (middle) and HTBD (right). It fails to detect backdoor samples crafted by SIG~and~HTBD.} 
    \label{fig:Strip}
    \vspace{-0.5em}
\end{figure*}

\textbf{SIG on GTSRB.} We run UltraClean with different removal thresholds ($\beta$ changes from 0 $\sim$ 0.3) on the entire training dataset and present the results in Table~\ref{tab:entire dataset}. In the experiment, 148 poisoned samples (3\% of the total training samples) are injected into the target class 5. ASR is evaluated at signal strength level $\Delta = 80$. UltraClean achieves \textgreater90\% ASR reduction while maintaining a high test accuracy, indicating the effectiveness of UltraClean on the entire training dataset. On the other hand, STRIP fails to detect the poisoned samples. The entropy distribution of poisoned samples is mostly overlapped with benign samples, as shown in Figure~\ref{fig:Strip} (left).

\textbf{LCBD on CIFAR-10.} We then evaluate UltraClean on the entire CIFAR-10 dataset. In this case, since AE ($\ell_2$, $\epsilon=1200$) has the best pre-clean ASR, we present the results under this setting. As shown in Table~\ref{tab:entire dataset}, UltraClean successfully thwarts the backdoor by diminishing the ASR to 1.59\%, reaching up to 97.40\% detection rate and 98.40\% ASR reduction. Meanwhile, even with the maximum removal threshold, the post-clean model accuracy only drops by $\sim$2.7\%. Note that STRIP also performs well on defending against LCBD. According to Figure~\ref{fig:Strip} (middle), the entropy of poisoned samples gathers around zero while the entropy of clean samples disperses in a wide range.

\textbf{HTBD on ImageNet.} The entire ImageNet dataset has more than one million training images. Precisely detecting poisoned samples from such a large-scale dataset is an extremely challenging task. However, UltraClean still acquires considerable success when facing such a complex dataset. In the experiment, we reproduce the attack in the original paper by injecting only 400 poisoned samples ($\sim$0.04\% of total training samples) into the target ``French bulldog'' class. The results of UltraClean and STRIP are presented in Table~\ref{tab:entire dataset} and Figure~\ref{fig:Strip} (right). It can be observed that UltraClean is particularly effective against HTBD, while STRIP is unsuccessful in detecting the poisoned samples. With only a 5\% removal threshold, UltraClean achieves a nearly 90\% detection rate while maintaining a similar level of model accuracy.

\begin{table}[htbp]
\setlength{\tabcolsep}{8pt}
\resizebox{\linewidth}{!}{
\begin{tabular}{c|cccc}
\toprule
\begin{tabular}[c]{@{}c@{}}\textbf{Removal}\\ \textbf{Threshold}\end{tabular} & \begin{tabular}[c]{@{}c@{}} \textbf{BDR}\end{tabular} & \begin{tabular}[c]{@{}c@{}} \textbf{Acc.}\end{tabular}     & \textbf{ASR}     & \begin{tabular}[c]{@{}c@{}}\textbf{ASR($\blacktriangledown$)}\\ \textbf{Reduction}\end{tabular} \\ \hline
\textbf{}  & \multicolumn{4}{c}{\textbf{SIG on GTSRB}}             \\ \cline{2-5}
\textbf{0.00}                                                          & 0.00\%                                                                                                                   & 95.49\% & 51.53\% & -                                                       \\
    \textbf{0.10}                                                          & 9.45\%                                                                                                                  & 96.52\% & 40.08\% & \textbf{22.22\%}                                                  \\
\textbf{0.20}                                                          & 25.00\%                                                                                                                 & 93.40\% & 10.69\% & \textbf{79.25\% }                                                 \\
\textbf{0.30}                                                           & 39.86\%                                                                                                                 & 88.54\% & 1.91\%  & \textbf{96.29\% }                                                \\
\hline   
& \multicolumn{4}{c}{\textbf{LCBD on CIFAR-10}}         \\ \cline{2-5} 
\textbf{0.00}                                                             & 0.00\%                                                                                                                           & 87.73\%                                                      & 99.98\%      & -                                                                  \\
\textbf{0.10}                                                             & 88.80\%                                                                                                                           & 87.29\%                                                             & 74.95\%             &  \textbf{25.00\%}                                                                  \\
\textbf{0.20}                                                             & 94.65\%                                                                                                                            & 86.66\%                                                             &  52.94\%            & \textbf{47.49\%}                                                                   \\
\textbf{0.30}                                                             & 97.40\%                                                                                                                            &  85.00\%                                                            & 1.59\%             & \textbf{98.40\%}                                                                   \\ \hline
& \multicolumn{4}{c}{\textbf{HTBD on ImageNet}}         \\ \cline{2-5} 
\textbf{0.00} & 0.00\%                                                                                                                                & 50.27\%                                                      & 48.00\%         & -                                                                  \\
\textbf{0.01} & 57.25\%                                                                                                                            &  50.06\%                                                            & 9.20\%             &  \textbf{80.83\%}                                                                  \\
\textbf{0.02} & 68.75\%                                                                                                                              & 50.26\%                                                         &  1.60\%            &    \textbf{96.00\%}                                                                \\
\textbf{0.03} & 78.00\%                                                                                                                              &   50.18\%                                                           &  0.20\%            &   \textbf{99.58\%}                                                                 \\
\textbf{0.04} & 83.50\%                                                                                                                            &    50.10\%                                                          &  0.00\%            &    \textbf{100.00\%}                                                                \\\bottomrule
\end{tabular}}
\caption{UltraClean performance on entire dataset.}
\label{tab:entire dataset}
\vspace{-0.5em}
\end{table}


\subsection{Performance on Clean Datasets}
{UltraClean is designed to detect and mitigate backdoor attacks when poisoned samples are injected into training datasets. However, in real-world scenarios, users typically do not know whether the training dataset is poisoned. Therefore, we also study the performance of UltraClean (removal threshold = 0.3) on clean dataset and dataset with Gaussian noise. The results are summarized in Table~\ref{Gaussian_noise}. We find that none of the models reveals a malicious backdoor behavior after training on these two datasets, and applying UltraClean does not affect the inference accuracy, which indicates that it is safe to apply UltraClean under any circumstance.}
\vspace{-0.5em}
\begin{table}[htbp]
\setlength{\tabcolsep}{8pt}
\resizebox{\linewidth}{!}{

\begin{tabular}{c|cc|cc}
\hline
\multirow{2}{*}{\textbf{\begin{tabular}[c]{@{}c@{}}Threshold\\  ($\boldsymbol\beta$ = 0.30)\end{tabular}}} & \multicolumn{2}{c|}{\textbf{Regular}} & \multicolumn{2}{c}{\textbf{UltraClean}} \\ \cline{2-5} 
                                                                                               & \multicolumn{1}{c}{Acc.}    & ASR   & \multicolumn{1}{c}{Acc.}      & ASR     \\ \hline
\textbf{Clean}                                                                                 & \multicolumn{1}{c}{85.97\%} & 0.0\% & \multicolumn{1}{c}{85.89\%}   & 0.0\%   \\ 
\textbf{Gaussian Noise}                                                                        & \multicolumn{1}{c}{85.03\%} & 0.8\% & \multicolumn{1}{c}{85.49\%}   & 0.0\%   \\ \hline
\end{tabular}}
\caption{Performance of UltraClean on clean dataset and Gaussian noise dataset}
\label{Gaussian_noise}
\vspace{-1em}
\end{table}


\subsection{Direct Training on Denoised Datasets}
{We also empirically evaluate if we can train backdoor-free models directly on denoised datasets. We apply non-local mean and local median denoising functions to the poisoned dataset crafted by BadNets. As shown in Table~\ref{direct_train_denoised_variants}, directly training on the denoised dataset fails to mitigate backdoor in the model. These experimental results further prove that simply applying image denoising functions is ineffective in cleansing the poisoned dataset, which aligns with the observation in Figure~\ref{fig:Explanation_UltraClean}.} 

\begin{table}[hthp]
\setlength{\tabcolsep}{24pt}
\resizebox{\linewidth}{!}{
\begin{tabular}{c|cc}
\hline
\textbf{}        & \textbf{Acc.} & \textbf{ASR} \\ \hline
\textbf{Regular} & 85.27\%       & 100.00\%     \\
\textbf{Median}  & 83.30\%       & 89.75\%      \\
\textbf{Mean}    & 84.16\%       & 90.00\%      \\ \hline
\end{tabular}}
\caption{Comparison of accuracy and ASR between training on regular poisoned dataset and denoised poisoned dataset (BadNets).}
\label{direct_train_denoised_variants}
\vspace{-0.8em}
\end{table}

\subsection{Robustness against Adaptive Attacks}
We also assess the performance of UltraClean against adaptive attacks to further demonstrate its capability. We consider a scenario where the adversary attempts to reduce the noise by altering adversarial perturbation to evade the defense of UltraClean. Here, as an example, we showcase the results of LCBD in Table~\ref{adaptive_LCBD} by lowering the interpolation ratio to make the attack more stealthy. As shown in the table, UltraClean achieves consistent superior performance against all adaptive attacks. A more comprehensive study is presented in the supplementary materials.

\begin{table}[htbp]
\centering
\setlength{\tabcolsep}{12pt}
\resizebox{\linewidth}{!}{
\begin{tabular}{c|ccc}
\hline
\textbf{Attack Type} & \textbf{Acc.} & \textbf{ASR}     & \textbf{BDR}     \\ \hline
\textbf{}            & \multicolumn{3}{c}{\textbf{UltraClean (UC)}}        \\ \cline{2-4} 
\textbf{GAN ($\uptau$ = 0.0)}         & 86.70\%       & \textbf{21.56\%} & \textbf{72.70\%} \\
\textbf{GAN ($\uptau$ = 0.1)}         & 86.82\%       & \textbf{47.59\%} & \textbf{70.95\%} \\
\textbf{GAN ($\uptau$ = 0.2)}         & 87.08\%       & \textbf{25.01\%} & \textbf{78.00\%} \\\hline
\textbf{AE ($\ell_2$, $\epsilon$ = 300)}          & 87.05\%       & \textbf{25.50\%} & \textbf{75.55\%} \\
\textbf{AE ($\ell_2$, $\epsilon$ = 600)}          & 86.77\%       & \textbf{27.74\%} & \textbf{88.65\%} \\ \hline
\textbf{AE ($\ell_\infty$, $\epsilon$ = 8)}          & 87.71\%       & \textbf{23.16\%} & \textbf{75.80\%} \\
\textbf{AE ($\ell_\infty$, $\epsilon$ = 16)}          & 86.81\%       & \textbf{35.16\%} & \textbf{89.40\%} \\ \hline
\end{tabular}}
\caption{Robustness of UltraClean against adaptive LCBD attacks on CIFAR-10.}
\label{adaptive_LCBD}
\vspace{-1em}
\end{table}

\section{Conclusion}
{This paper presented a general defensive framework, UltraClean, against various backdoor attacks. It effectively differentiates poisoned and benign samples and significantly reduces the backdoor attack success rate. Comprehensive experiments and analysis validate the effectiveness of UltraClean in defending against both dirty-label and clean-label attacks.}

\section*{ACKNOWLEDGMENT}
This work is supported in part by the National Science Foundation (NSF) under grant numbers 2426299, 2413046, and 2532588.

{
    \small
    \bibliographystyle{ieeenat_fullname}
    \bibliography{main}

@String(CVPR= {IEEE Conf. Comput. Vis. Pattern Recog.})

@String(ICCV= {Int. Conf. Comput. Vis.})

@String(ECCV= {Eur. Conf. Comput. Vis.})

@String(ICIP = {IEEE Int. Conf. Image Process.})

@String(ICLR = {Int. Conf. Learn. Represent.})

@String(IJCAI = {IJCAI})

@String(AAAI = {AAAI})

@String(CVPR  = {CVPR})

@String(ICCV  = {ICCV})

@String(ECCV  = {ECCV})

@String(ICIP  = {ICIP})

@String(ICLR  = {ICLR})

@article{gu2017badnets,
  title={Badnets: Identifying vulnerabilities in the machine learning model supply chain},
  author={Gu, Tianyu and Dolan-Gavitt, Brendan and Garg, Siddharth},
  journal={arXiv preprint arXiv:1708.06733},
  year={2017}
}

@article{DBLP:journals/corr/abs-1712-05526,
  author    = {Xinyun Chen and
               Chang Liu and
               Bo Li and
               Kimberly Lu and
               Dawn Song},
  title     = {Targeted Backdoor Attacks on Deep Learning Systems Using Data Poisoning},
  journal   = {CoRR},
  volume    = {abs/1712.05526},
  year      = {2017}
}

@inproceedings{DBLP:conf/codaspy/ZhongLSZ020,
  author    = {Haoti Zhong and
               Cong Liao and
               Anna Cinzia Squicciarini and
               Sencun Zhu and
               David J. Miller},
  title     = {Backdoor Embedding in Convolutional Neural Network Models via Invisible
               Perturbation},
  booktitle = {{CODASPY} '20: Tenth {ACM} Conference on Data and Application Security
               and Privacy},
  pages     = {97--108},
  publisher = {{ACM}},
  year      = {2020}
}

@inproceedings{DBLP:conf/iclr/NguyenT21,
  author    = {Tuan Anh Nguyen and
               Anh Tuan Tran},
  title     = {WaNet - Imperceptible Warping-based Backdoor Attack},
  booktitle = {9th International Conference on Learning Representations, {ICLR}},
  year      = {2021}
}

@inproceedings{DBLP:conf/cvpr/Moosavi-Dezfooli17,
  author    = {Seyed{-}Mohsen Moosavi{-}Dezfooli and
               Alhussein Fawzi and
               Omar Fawzi and
               Pascal Frossard},
  title     = {Universal Adversarial Perturbations},
  booktitle = {2017 {IEEE} Conference on Computer Vision and Pattern Recognition,
               {CVPR}},
  pages     = {86--94},
  publisher = {{IEEE} Computer Society},
  year      = {2017}
}

@article{DBLP:journals/tdsc/LiXZZZ21,
  author    = {Shaofeng Li and
               Minhui Xue and
               Benjamin Zi Hao Zhao and
               Haojin Zhu and
               Xinpeng Zhang},
  title     = {Invisible Backdoor Attacks on Deep Neural Networks Via Steganography
               and Regularization},
  journal   = {{IEEE} Trans. Dependable Secur. Comput.},
  volume    = {18},
  number    = {5},
  pages     = {2088--2105},
  year      = {2021}
}

@inproceedings{li2021invisible,
  title={Invisible Backdoor Attack With Sample-Specific Triggers},
  author={Li, Yuezun and Li, Yiming and Wu, Baoyuan and Li, Longkang and He, Ran and Lyu, Siwei},
  booktitle={Proceedings of the IEEE/CVF International Conference on Computer Vision},
  pages={16463--16472},
  year={2021}
}

@inproceedings{DBLP:conf/aaai/0005LMZ21,
  author    = {Siyuan Cheng and
               Yingqi Liu and
               Shiqing Ma and
               Xiangyu Zhang},
  title     = {Deep Feature Space Trojan Attack of Neural Networks by Controlled
               Detoxification},
  booktitle = {Thirty-Fifth {AAAI} Conference on Artificial Intelligence, {AAAI}
               2021, Thirty-Third Conference on Innovative Applications of Artificial
               Intelligence},
  pages     = {1148--1156},
  publisher = {{AAAI} Press},
  year      = {2021}
}

@inproceedings{DBLP:conf/icip/BarniKT19,
  author    = {Mauro Barni and
               Kassem Kallas and
               Benedetta Tondi},
  title     = {A New Backdoor Attack in {CNNS} by Training Set Corruption Without
               Label Poisoning},
  booktitle = {2019 {IEEE} International Conference on Image Processing, {ICIP}},
  pages     = {101--105},
  publisher = {{IEEE}},
  year      = {2019}
}

@inproceedings{DBLP:conf/eccv/LiuM0020,
  author    = {Yunfei Liu and
               Xingjun Ma and
               James Bailey and
               Feng Lu},
  title     = {Reflection Backdoor: {A} Natural Backdoor Attack on Deep Neural Networks},
  booktitle = {Computer Vision - {ECCV} 2020 - 16th European Conference, Glasgow,
               UK, August 23-28, 2020, Proceedings, Part {X}},
  series    = {Lecture Notes in Computer Science},
  volume    = {12355},
  pages     = {182--199},
  publisher = {Springer},
  year      = {2020}
}

@inproceedings{DBLP:conf/aaai/SahaSP20,
  author    = {Aniruddha Saha and
               Akshayvarun Subramanya and
               Hamed Pirsiavash},
  title     = {Hidden Trigger Backdoor Attacks},
  booktitle = {The Thirty-Fourth {AAAI} Conference on Artificial Intelligence, {AAAI}
               2020},
  pages     = {11957--11965},
  publisher = {{AAAI} Press},
  year      = {2020}
}

@article{DBLP:journals/corr/abs-1912-02771,
  author    = {Alexander Turner and
               Dimitris Tsipras and
               Aleksander Madry},
  title     = {Label-Consistent Backdoor Attacks},
  journal   = {CoRR},
  volume    = {abs/1912.02771},
  year      = {2019}
}

@inproceedings{DBLP:conf/sp/QuiringR20,
  author    = {Erwin Quiring and
               Konrad Rieck},
  title     = {Backdooring and Poisoning Neural Networks with Image-Scaling Attacks},
  booktitle = {2020 {IEEE} Security and Privacy Workshops, {SP} Workshops},
  pages     = {41--47},
  publisher = {{IEEE}},
  year      = {2020}
}

@inproceedings{DBLP:conf/iclr/GeipingFHCT0G21,
  author    = {Jonas Geiping and
               Liam H. Fowl and
               W. Ronny Huang and
               Wojciech Czaja and
               Gavin Taylor and
               Michael Moeller and
               Tom Goldstein},
  title     = {Witches' Brew: Industrial Scale Data Poisoning via Gradient Matching},
  booktitle = {9th International Conference on Learning Representations, {ICLR}},
  year      = {2021}
}

@inproceedings{DBLP:journals/corr/GoodfellowSS14,
  author    = {Ian J. Goodfellow and
               Jonathon Shlens and
               Christian Szegedy},
  editor    = {Yoshua Bengio and
               Yann LeCun},
  title     = {Explaining and Harnessing Adversarial Examples},
  booktitle = {3rd International Conference on Learning Representations, {ICLR}},
  year      = {2015}
}

@inproceedings{DBLP:conf/nips/Tran0M18,
  author    = {Brandon Tran and
               Jerry Li and
               Aleksander Madry},
  title     = {Spectral Signatures in Backdoor Attacks},
  booktitle = {NeurIPS},
  pages     = {8011--8021},
  year      = {2018}
}

@article{DBLP:journals/corr/abs-2012-10544,
  author    = {Micah Goldblum and
               Dimitris Tsipras and
               Chulin Xie and
               Xinyun Chen and
               Avi Schwarzschild and
               Dawn Song and
               Aleksander Madry and
               Bo Li and
               Tom Goldstein},
  title     = {Dataset Security for Machine Learning: Data Poisoning, Backdoor Attacks,
               and Defenses},
  journal   = {CoRR},
  volume    = {abs/2012.10544},
  year      = {2020}
}

@article{DBLP:journals/corr/abs-2007-08745,
  author    = {Yiming Li and
               Baoyuan Wu and
               Yong Jiang and
               Zhifeng Li and
               Shu{-}Tao Xia},
  title     = {Backdoor Learning: {A} Survey},
  journal   = {CoRR},
  volume    = {abs/2007.08745},
  year      = {2020}
}

@inproceedings{DBLP:conf/iclr/DuJS20,
  author    = {Min Du and
               Ruoxi Jia and
               Dawn Song},
  title     = {Robust anomaly detection and backdoor attack detection via differential
               privacy},
  booktitle = {8th International Conference on Learning Representations, {ICLR}},
  year      = {2020}
}

@article{DBLP:journals/corr/abs-2002-11497,
  author    = {Sanghyun Hong and
               Varun Chandrasekaran and
               Yigitcan Kaya and
               Tudor Dumitras and
               Nicolas Papernot},
  title     = {On the Effectiveness of Mitigating Data Poisoning Attacks with Gradient
               Shaping},
  journal   = {CoRR},
  volume    = {abs/2002.11497},
  year      = {2020}
}

@inproceedings{DBLP:conf/aaai/ChenCBLELMS19,
  author    = {Bryant Chen and
               Wilka Carvalho and
               Nathalie Baracaldo and
               Heiko Ludwig and
               Benjamin Edwards and
               Taesung Lee and
               Ian M. Molloy and
               Biplav Srivastava},
  title     = {Detecting Backdoor Attacks on Deep Neural Networks by Activation Clustering},
  booktitle = {Workshop on Artificial Intelligence Safety 2019 co-located with the
               Thirty-Third {AAAI} Conference on Artificial Intelligence {AAAI}},
  year      = {2019}
}

@inproceedings{DBLP:conf/sp/ChouTP20,
  author    = {Edward Chou and
               Florian Tram{\`{e}}r and
               Giancarlo Pellegrino},
  title     = {SentiNet: Detecting Localized Universal Attacks Against Deep Learning
               Systems},
  booktitle = {2020 {IEEE} Security and Privacy Workshops, {SP} Workshops},
  pages     = {48--54},
  publisher = {{IEEE}},
  year      = {2020}
}

@inproceedings{DBLP:conf/uss/Tang0TZ21,
  author    = {Di Tang and
               XiaoFeng Wang and
               Haixu Tang and
               Kehuan Zhang},
  title     = {Demon in the Variant: Statistical Analysis of DNNs for Robust Backdoor
               Contamination Detection},
  booktitle = {30th {USENIX} Security Symposium, {USENIX} Security},
  pages     = {1541--1558},
  publisher = {{USENIX} Association},
  year      = {2021}
}

@inproceedings{DBLP:conf/acsac/GaoXW0RN19,
  author    = {Yansong Gao and
               Change Xu and
               Derui Wang and
               Shiping Chen and
               Damith Chinthana Ranasinghe and
               Surya Nepal},
  title     = {{STRIP:} a defence against trojan attacks on deep neural networks},
  booktitle = {Proceedings of the 35th Annual Computer Security Applications Conference,
               {ACSAC}},
  pages     = {113--125},
  publisher = {{ACM}},
  year      = {2019}
}

@inproceedings{DBLP:conf/raid/0017DG18,
  author    = {Kang Liu and
               Brendan Dolan{-}Gavitt and
               Siddharth Garg},
  title     = {Fine-Pruning: Defending Against Backdooring Attacks on Deep Neural
               Networks},
  booktitle = {Research in Attacks, Intrusions, and Defenses - 21st International
               Symposium, {RAID} Proceedings},
  series    = {Lecture Notes in Computer Science},
  volume    = {11050},
  pages     = {273--294},
  publisher = {Springer},
  year      = {2018}
}

@inproceedings{DBLP:conf/iclr/LiLKLLM21,
  author    = {Yige Li and
               Xixiang Lyu and
               Nodens Koren and
               Lingjuan Lyu and
               Bo Li and
               Xingjun Ma},
  title     = {Neural Attention Distillation: Erasing Backdoor Triggers from Deep
               Neural Networks},
  booktitle = {9th International Conference on Learning Representations, {ICLR}},
  year      = {2021}
}

@inproceedings{DBLP:conf/iclr/ZhaoCDRL20,
  author    = {Pu Zhao and
               Pin{-}Yu Chen and
               Payel Das and
               Karthikeyan Natesan Ramamurthy and
               Xue Lin},
  title     = {Bridging Mode Connectivity in Loss Landscapes and Adversarial Robustness},
  booktitle = {8th International Conference on Learning Representations, {ICLR}},
  year      = {2020}
}

@article{DBLP:journals/corr/abs-1911-07399,
  author    = {Xijie Huang and
               Moustafa Alzantot and
               Mani B. Srivastava},
  title     = {NeuronInspect: Detecting Backdoors in Neural Networks via Output Explanations},
  journal   = {CoRR},
  volume    = {abs/1911.07399},
  year      = {2019}
}

@inproceedings{DBLP:conf/sp/XuWLBGL21,
  author    = {Xiaojun Xu and
               Qi Wang and
               Huichen Li and
               Nikita Borisov and
               Carl A. Gunter and
               Bo Li},
  title     = {Detecting {AI} Trojans Using Meta Neural Analysis},
  booktitle = {42nd {IEEE} Symposium on Security and Privacy, {SP}},
  pages     = {103--120},
  publisher = {{IEEE}},
  year      = {2021}
}

@inproceedings{DBLP:conf/eccv/HuangPJT20,
  author    = {Shanjiaoyang Huang and
               Weiqi Peng and
               Zhiwei Jia and
               Zhuowen Tu},
  title     = {One-Pixel Signature: Characterizing {CNN} Models for Backdoor Detection},
  booktitle = {Computer Vision - {ECCV} 2020 - 16th European Conference, Proceedings, Part {XXVII}},
  series    = {Lecture Notes in Computer Science},
  volume    = {12372},
  pages     = {326--341},
  publisher = {Springer},
  year      = {2020}
}

@inproceedings{DBLP:conf/eccv/WangZLCXW20,
  author    = {Ren Wang and
               Gaoyuan Zhang and
               Sijia Liu and
               Pin{-}Yu Chen and
               Jinjun Xiong and
               Meng Wang},
  title     = {Practical Detection of Trojan Neural Networks: Data-Limited and Data-Free
               Cases},
  booktitle = {Computer Vision - {ECCV} 2020 - 16th European Conference, Proceedings, Part {XXIII}},
  year      = {2020}
}

@inproceedings{DBLP:conf/cvpr/KolouriSPH20,
  author    = {Soheil Kolouri and
               Aniruddha Saha and
               Hamed Pirsiavash and
               Heiko Hoffmann},
  title     = {Universal Litmus Patterns: Revealing Backdoor Attacks in CNNs},
  booktitle = {2020 {IEEE/CVF} Conference on Computer Vision and Pattern Recognition,
               {CVPR}},
  pages     = {298--307},
  publisher = {Computer Vision Foundation / {IEEE}},
  year      = {2020}
}

@inproceedings{DBLP:conf/acsac/DoanAR20,
  author    = {Bao Gia Doan and
               Ehsan Abbasnejad and
               Damith C. Ranasinghe},
  title     = {Februus: Input Purification Defense Against Trojan Attacks on Deep
               Neural Network Systems},
  booktitle = {{ACSAC} '20: Annual Computer Security Applications Conference},
  pages     = {897--912},
  publisher = {{ACM}},
  year      = {2020}
}

@article{udeshi2019model,
  title={Model agnostic defence against backdoor attacks in machine learning},
  author={Udeshi, Sakshi and Peng, Shanshan and Woo, Gerald and Loh, Lionell and Rawshan, Louth and Chattopadhyay, Sudipta},
  journal={arXiv preprint arXiv:1908.02203},
  year={2019}
}

@inproceedings{qiu2021deepsweep,
  title={Deepsweep: An evaluation framework for mitigating dnn backdoor attacks using data augmentation},
  author={Qiu, Han and Zeng, Yi and Guo, Shangwei and Zhang, Tianwei and Qiu, Meikang and Thuraisingham, Bhavani},
  booktitle={Proceedings of the 2021 ACM Asia Conference on Computer and Communications Security},
  pages={363--377},
  year={2021}
}

@inproceedings{javaheripi2020cleann,
  title={Cleann: Accelerated trojan shield for embedded neural networks},
  author={Javaheripi, Mojan and Samragh, Mohammad and Fields, Gregory and Javidi, Tara and Koushanfar, Farinaz},
  booktitle={2020 IEEE/ACM International Conference On Computer Aided Design (ICCAD)},
  pages={1--9},
  year={2020},
  organization={IEEE}
}

@article{jin2020unified,
  title={A unified framework for analyzing and detecting malicious examples of dnn models},
  author={Jin, Kaidi and Zhang, Tianwei and Shen, Chao and Chen, Yufei and Fan, Ming and Lin, Chenhao and Liu, Ting},
  journal={arXiv preprint arXiv:2006.14871},
  year={2020}
}

@inproceedings{wang2019neural,
  title={Neural cleanse: Identifying and mitigating backdoor attacks in neural networks},
  author={Wang, Bolun and Yao, Yuanshun and Shan, Shawn and Li, Huiying and Viswanath, Bimal and Zheng, Haitao and Zhao, Ben Y},
  booktitle={2019 IEEE Symposium on Security and Privacy (SP)},
  pages={707--723},
  year={2019}
}

@inproceedings{chen2019deepinspect,
  title={DeepInspect: A Black-box Trojan Detection and Mitigation Framework for Deep Neural Networks.},
  author={Chen, Huili and Fu, Cheng and Zhao, Jishen and Koushanfar, Farinaz},
  booktitle={IJCAI},
  pages={4658--4664},
  year={2019}
}

@inproceedings{DBLP:conf/nips/QiaoYL19,
  author    = {Ximing Qiao and
               Yukun Yang and
               Hai Li},
  title     = {Defending Neural Backdoors via Generative Distribution Modeling},
  booktitle = {Advances in Neural Information Processing Systems 32, {NeurIPS}},
  pages     = {14004--14013},
  year      = {2019}
}

@inproceedings{DBLP:conf/icdm/GuoWXXDS20,
  author    = {Wenbo Guo and
               Lun Wang and
               Yan Xu and
               Xinyu Xing and
               Min Du and
               Dawn Song},
  title     = {Towards Inspecting and Eliminating Trojan Backdoors in Deep Neural
               Networks},
  booktitle = {20th {ICDM}},
  year      = {2020}
}

@inproceedings{DBLP:conf/mm/ZhuNWXW20,
  author    = {Liuwan Zhu and
               Rui Ning and
               Cong Wang and
               Chunsheng Xin and
               Hongyi Wu},
  title     = {GangSweep: Sweep out Neural Backdoors by {GAN}},
  booktitle = {{MM} '20: The 28th {ACM} International Conference on Multimedia,},
  pages     = {3173--3181},
  publisher = {{ACM}},
  year      = {2020}
}

@inproceedings{he2016deep,
  title={Deep residual learning for image recognition},
  author={He, Kaiming and Zhang, Xiangyu and Ren, Shaoqing and Sun, Jian},
  booktitle={Proceedings of the IEEE conference on computer vision and pattern recognition},
  pages={770--778},
  year={2016}
}

@inproceedings{DBLP:conf/iclr/DosovitskiyB0WZ21,
  author    = {Alexey Dosovitskiy and
               Lucas Beyer and
               Alexander Kolesnikov and
               Dirk Weissenborn and
               Xiaohua Zhai and
               Thomas Unterthiner and
               Mostafa Dehghani and
               Matthias Minderer and
               Georg Heigold and
               Sylvain Gelly and
               Jakob Uszkoreit and
               Neil Houlsby},
  title     = {An Image is Worth 16x16 Words: Transformers for Image Recognition
               at Scale},
  booktitle = {9th International Conference on Learning Representations, {ICLR}},

  year      = {2021}
}

@inproceedings{DBLP:conf/naacl/DevlinCLT19,
  author    = {Jacob Devlin and
               Ming{-}Wei Chang and
               Kenton Lee and
               Kristina Toutanova},
  title     = {{BERT:} Pre-training of Deep Bidirectional Transformers for Language
               Understanding},
  booktitle = {Proceedings of the 2019 NAACL-HLT},
  pages     = {4171--4186},
  publisher = {Association for Computational Linguistics},
  year      = {2019}
}

@inproceedings{DBLP:conf/nips/BrownMRSKDNSSAA20,
  author    = {Tom B. Brown and
               Benjamin Mann and
               Nick Ryder and
               Melanie Subbiah and
               Jared Kaplan and
               Prafulla Dhariwal and
               Arvind Neelakantan and
               Pranav Shyam and
               Girish Sastry and
               Amanda Askell and
               Sandhini Agarwal and
               Ariel Herbert{-}Voss and
               Gretchen Krueger and
               Tom Henighan and
               Rewon Child and
               Aditya Ramesh and
               Daniel M. Ziegler and
               Jeffrey Wu and
               Clemens Winter and
               Christopher Hesse and
               Mark Chen and
               Eric Sigler and
               Mateusz Litwin and
               Scott Gray and
               Benjamin Chess and
               Jack Clark and
               Christopher Berner and
               Sam McCandlish and
               Alec Radford and
               Ilya Sutskever and
               Dario Amodei},
  title     = {Language Models are Few-Shot Learners},
  booktitle = {Advances in Neural Information Processing Systems 33: Annual Conference
               on Neural Information Processing Systems {NeurIPS}},
  year      = {2020}
}

@article{DBLP:journals/nature/SilverHMGSDSAPL16,
  author    = {David Silver and
               Aja Huang and
               Chris J. Maddison and
               Arthur Guez and
               Laurent Sifre and
               George van den Driessche and
               Julian Schrittwieser and
               Ioannis Antonoglou and
               Vedavyas Panneershelvam and
               Marc Lanctot and
               Sander Dieleman and
               Dominik Grewe and
               John Nham and
               Nal Kalchbrenner and
               Ilya Sutskever and
               Timothy P. Lillicrap and
               Madeleine Leach and
               Koray Kavukcuoglu and
               Thore Graepel and
               Demis Hassabis},
  title     = {Mastering the game of Go with deep neural networks and tree search},
  journal   = {Nat.},
  volume    = {529},
  number    = {7587},
  pages     = {484--489},
  year      = {2016}
}

@article{DBLP:journals/corr/BojarskiTDFFGJM16,
  author    = {Mariusz Bojarski and
               Davide Del Testa and
               Daniel Dworakowski and
               Bernhard Firner and
               Beat Flepp and
               Prasoon Goyal and
               Lawrence D. Jackel and
               Mathew Monfort and
               Urs Muller and
               Jiakai Zhang and
               Xin Zhang and
               Jake Zhao and
               Karol Zieba},
  title     = {End to End Learning for Self-Driving Cars},
  journal   = {CoRR},
  volume    = {abs/1604.07316},
  year      = {2016}
}

@article{DBLP:journals/ijon/WangD21a,
  author    = {Mei Wang and
               Weihong Deng},
  title     = {Deep face recognition: {A} survey},
  journal   = {Neurocomputing},
  volume    = {429},
  pages     = {215--244},
  year      = {2021}
}

@inproceedings{DBLP:conf/ndss/Xu0Q18,
  author    = {Weilin Xu and
               David Evans and
               Yanjun Qi},
  title     = {Feature Squeezing: Detecting Adversarial Examples in Deep Neural Networks},
  booktitle = {25th Annual Network and Distributed System Security Symposium, {NDSS}},
  publisher = {The Internet Society},
  year      = {2018}
}

@inproceedings{DBLP:conf/cvpr/XieWMYH19,
  author    = {Cihang Xie and
               Yuxin Wu and
               Laurens van der Maaten and
               Alan L. Yuille and
               Kaiming He},
  title     = {Feature Denoising for Improving Adversarial Robustness},
  booktitle = {{IEEE} Conference on Computer Vision and Pattern Recognition, {CVPR}},
  pages     = {501--509},
  publisher = {Computer Vision Foundation / {IEEE}},
  year      = {2019}
}

@inproceedings{DBLP:conf/icml/BiggioNL12,
  author    = {Battista Biggio and
               Blaine Nelson and
               Pavel Laskov},
  title     = {Poisoning Attacks against Support Vector Machines},
  booktitle = {Proceedings of the 29th International Conference on Machine Learning,
               {ICML}},
  publisher = {icml.cc / Omnipress},
  year      = {2012}
}

@inproceedings{DBLP:conf/cvpr/BuadesCM05,
  author    = {Antoni Buades and
               Bartomeu Coll and
               Jean{-}Michel Morel},
  title     = {A Non-Local Algorithm for Image Denoising},
  booktitle = {2005 {IEEE} Computer Society Conference on Computer Vision and Pattern
               Recognition {CVPR}},
  pages     = {60--65},
  publisher = {{IEEE} Computer Society},
  year      = {2005}
}

@misc{tensorflow2015-whitepaper,
title={ {TensorFlow}: Large-Scale Machine Learning on Heterogeneous Systems},
author={
    Mart\'{\i}n~Abadi and
    Ashish~Agarwal and
    Paul~Barham and
    Eugene~Brevdo and
    Zhifeng~Chen and
    Craig~Citro and
    Greg~S.~Corrado and
    Andy~Davis and
    Jeffrey~Dean and
    Matthieu~Devin and
    Sanjay~Ghemawat and
    Ian~Goodfellow and
    Andrew~Harp and
    Geoffrey~Irving and
    Michael~Isard and
    Yangqing Jia and
    Rafal~Jozefowicz and
    Lukasz~Kaiser and
    Manjunath~Kudlur and
    Josh~Levenberg and
    Dandelion~Man\'{e} and
    Rajat~Monga and
    Sherry~Moore and
    Derek~Murray and
    Chris~Olah and
    Mike~Schuster and
    Jonathon~Shlens and
    Benoit~Steiner and
    Ilya~Sutskever and
    Kunal~Talwar and
    Paul~Tucker and
    Vincent~Vanhoucke and
    Vijay~Vasudevan and
    Fernanda~Vi\'{e}gas and
    Oriol~Vinyals and
    Pete~Warden and
    Martin~Wattenberg and
    Martin~Wicke and
    Yuan~Yu and
    Xiaoqiang~Zheng},
  year={2015},
}

@article{paszke2019pytorch,
  title={Pytorch: An imperative style, high-performance deep learning library},
  author={Paszke, Adam and Gross, Sam and Massa, Francisco and Lerer, Adam and Bradbury, James and Chanan, Gregory and Killeen, Trevor and Lin, Zeming and Gimelshein, Natalia and Antiga, Luca and others},
  journal={Advances in neural information processing systems},
  volume={32},
  pages={8026--8037},
  year={2019}
}

@inproceedings{deng2009imagenet,
  title={Imagenet: A large-scale hierarchical image database},
  author={Deng, Jia and Dong, Wei and Socher, Richard and Li, Li-Jia and Li, Kai and Fei-Fei, Li},
  booktitle={2009 IEEE conference on computer vision and pattern recognition},
  pages={248--255},
  year={2009},
  organization={Ieee}
}

@inproceedings{stallkamp2011german,
  title={The German traffic sign recognition benchmark: a multi-class classification competition},
  author={Stallkamp, Johannes and Schlipsing, Marc and Salmen, Jan and Igel, Christian},
  booktitle={The 2011 international joint conference on neural networks},
  pages={1453--1460},
  year={2011},
  organization={IEEE}
}

@article{DBLP:journals/corr/abs-2110-11571,
  author    = {Yige Li and
               Xixiang Lyu and
               Nodens Koren and
               Lingjuan Lyu and
               Bo Li and
               Xingjun Ma},
  title     = {Anti-Backdoor Learning: Training Clean Models on Poisoned Data},
  journal   = {CoRR},
  volume    = {abs/2110.11571},
  year      = {2021}
}

@inproceedings{DBLP:conf/iclr/LinGH19,
  author    = {Ji Lin and
               Chuang Gan and
               Song Han},
  title     = {Defensive Quantization: When Efficiency Meets Robustness},
  booktitle = {7th International Conference on Learning Representations, {ICLR}},
  year      = {2019}
}

@inproceedings{Trojannn,
  author    = {Yingqi Liu and
               Shiqing Ma and
               Yousra Aafer and
               Wen-Chuan Lee and
               Juan Zhai and
               Weihang Wang and
               Xiangyu Zhang},
  title     = {Trojaning Attack on Neural Networks},
  booktitle = {25th Annual Network and Distributed System Security Symposium, {NDSS}},
  publisher = {The Internet Society},
  year      = {2018},
}

@article{DBLP:journals/corr/abs-1803-00992,
  author    = {Andrea Paudice and
               Luis Mu{\~{n}}oz{-}Gonz{\'{a}}lez and
               Emil C. Lupu},
  title     = {Label Sanitization against Label Flipping Poisoning Attacks},
  journal   = {CoRR},
  volume    = {abs/1803.00992},
  year      = {2018}
}

@article{DBLP:journals/corr/abs-2202-03423,
  author    = {Kunzhe Huang and
               Yiming Li and
               Baoyuan Wu and
               Zhan Qin and
               Kui Ren},
  title     = {Backdoor Defense via Decoupling the Training Process},
  journal   = {CoRR},
  volume    = {abs/2202.03423},
  year      = {2022}
}

@inproceedings{DBLP:conf/iccv/ZengPMJ21,
  author    = {Yi Zeng and
               Won Park and
               Z. Morley Mao and
               Ruoxi Jia},
  title     = {Rethinking the Backdoor Attacks' Triggers: {A} Frequency Perspective},
  booktitle = {2021 {IEEE/CVF} International Conference on Computer Vision, {ICCV}},
  pages     = {16453--16461},
  year      = {2021}}

@article{DBLP:journals/corr/abs-2104-11315,
  author    = {Jonathan Hayase and
               Weihao Kong and
               Raghav Somani and
               Sewoong Oh},
  title     = {{SPECTRE:} Defending Against Backdoor Attacks Using Robust Statistics},
  journal   = {CoRR},
  volume    = {abs/2104.11315},
  year      = {2021}
}

@inproceedings{DBLP:conf/iccv/DoanL0L21,
  author    = {Khoa Doan and
               Yingjie Lao and
               Weijie Zhao and
               Ping Li},
  title     = {{LIRA:} Learnable, Imperceptible and Robust Backdoor Attacks},
  booktitle = {2021 {IEEE/CVF} International Conference on Computer Vision, {ICCV}
               2021, Montreal, QC, Canada, October 10-17, 2021},
  pages     = {11946--11956},
  publisher = {{IEEE}},
  year      = {2021}
}

@inproceedings{DBLP:conf/nips/DoanLL21,
  author    = {Khoa Doan and
               Yingjie Lao and
               Ping Li},
  editor    = {Marc'Aurelio Ranzato and
               Alina Beygelzimer and
               Yann N. Dauphin and
               Percy Liang and
               Jennifer Wortman Vaughan},
  title     = {Backdoor Attack with Imperceptible Input and Latent Modification},
  booktitle = {Advances in Neural Information Processing Systems 34: Annual Conference
               on Neural Information Processing Systems 2021, NeurIPS 2021, December
               6-14, 2021, virtual},
  pages     = {18944--18957},
  year      = {2021}
}

@inproceedings{DBLP:conf/nips/NguyenT20,
  author    = {Tuan Anh Nguyen and
               Anh Tuan Tran},
  editor    = {Hugo Larochelle and
               Marc'Aurelio Ranzato and
               Raia Hadsell and
               Maria{-}Florina Balcan and
               Hsuan{-}Tien Lin},
  title     = {Input-Aware Dynamic Backdoor Attack},
  booktitle = {Advances in Neural Information Processing Systems 33: Annual Conference
               on Neural Information Processing Systems 2020, NeurIPS 2020, December
               6-12, 2020, virtual},
  year      = {2020}
}

@article{souri2022sleeper,
  title={Sleeper agent: Scalable hidden trigger backdoors for neural networks trained from scratch},
  author={Souri, Hossein and Fowl, Liam and Chellappa, Rama and Goldblum, Micah and Goldstein, Tom},
  journal={Advances in Neural Information Processing Systems},
  volume={35},
  pages={19165--19178},
  year={2022}
}

@article{wu2021adversarial,
  title={Adversarial neuron pruning purifies backdoored deep models},
  author={Wu, Dongxian and Wang, Yisen},
  journal={Advances in Neural Information Processing Systems},
  volume={34},
  pages={16913--16925},
  year={2021}
}

@inproceedings{zhao2022clpa,
  title={CLPA: Clean-label poisoning availability attacks using generative adversarial nets},
  author={Zhao, Bingyin and Lao, Yingjie},
  booktitle={Proceedings of the AAAI Conference on Artificial Intelligence},
  volume={36},
  number={8},
  pages={9162--9170},
  year={2022}
}

@inproceedings{zhao2022towards,
  title={Towards class-oriented poisoning attacks against neural networks},
  author={Zhao, Bingyin and Lao, Yingjie},
  booktitle={Proceedings of the IEEE/CVF Winter Conference on Applications of Computer Vision},
  pages={3741--3750},
  year={2022}
}

@inproceedings{zhao2018resilience,
  title={Resilience of pruned neural network against poisoning attack},
  author={Zhao, Bingyin and Lao, Yingjie},
  booktitle={2018 13th International Conference on Malicious and Unwanted Software (MALWARE)},
  pages={78--83},
  year={2018},
  organization={IEEE}
}

@inproceedings{DBLP:conf/cvpr/HanZCL0L25,
  author       = {Yuning Han and
                  Bingyin Zhao and
                  Rui Chu and
                  Feng Luo and
                  Biplab Sikdar and
                  Yingjie Lao},
  title        = {UIBDiffusion: Universal Imperceptible Backdoor Attack for Diffusion
                  Models},
  booktitle    = {{IEEE/CVF} Conference on Computer Vision and Pattern Recognition,
                  {CVPR} 2025, Nashville, TN, USA, June 11-15, 2025},
  pages        = {19186--19196},
  publisher    = {Computer Vision Foundation / {IEEE}},
  year         = {2025}
}
}

\newpage
\appendix


\section{Mechanisms of Different Poisons}
As discussed in Section 2 in the main paper, various backdoor attacks have been developed in the literature. In this paper, we examine several of the most representative attacks: BadNets~\cite{gu2017badnets}, blended injection attack~\cite{DBLP:journals/corr/abs-1712-05526}, Trojan~\cite{Trojannn}, sinusoidal signal backdoor~\cite{DBLP:conf/icip/BarniKT19}, label-consistent backdoor~\cite{DBLP:journals/corr/abs-1912-02771}, and hidden trigger backdoor~\cite{DBLP:conf/aaai/SahaSP20}. We elucidate the mechanisms of these attacks in this section.

\subsection{Dirty-Label Attacks}

\hspace{1.2em} \textbf{BadNets.} BadNets is the first backdoor attack against deep neural networks, which uses a pattern (e.g., flower, colored square, boom, etc.) as the backdoor trigger. Poisoned samples are patched with the backdoor trigger and assigned an incorrect target label. The poisons generation process are expressed in Equation~\ref{equ: badnets_generation}, where $\mathcal{T}$ is the backdoor trigger and $\mathcal{M}$ is the masking area to put the trigger.

\begin{equation}
\begin{aligned}
     x_p &= x_b + \mathcal{T} \odot \mathcal{M},\\
     y_p &= y_b \leftarrow y_t,
     \label{equ: badnets_generation}
\end{aligned}
\end{equation}

\textbf{Blended Injection Attack.} Blended crafts backdoor trigger using a fixed cartoon pattern (e.g., Hello Kitty) or a random pattern (e.g., random noise). Similar to BadNets, a target dirty-label is assigned to all poisoned samples. Equation~\ref{equ: blended_generation} depicts the generation process of Blended, where $\alpha$ is the blended ratio.

\begin{equation}
\begin{aligned}
    x_p = \prod^{\textbf{blend}}_{\alpha}(\mathcal{T},x_b) &= \alpha \odot \mathcal{T} + (1-\alpha) \odot x_b,\\
    y_p &= y_b \leftarrow y_t,
    \label{equ: blended_generation}
\end{aligned}
\end{equation}

\textbf{Trojan.} Trojan generates the backdoor trigger by assigning value to the input variables in a given trigger mask. The value assignment process aims to make some selected internal neurons (i.e., a neuron in the last layer denoting the target dirty-label) to achieve the maximum values. The neurons selection process can be mathematically expressed by Equation~\ref{equ: trojan_neuron_selection} where $L_{target}$ is the layer of target neurons, $L_{preceding}$ is the preceding layers, $W$ is the weights and $b$ is the biases between the layers.

\begin{equation}
\begin{aligned}
    L_{target} = L_{preceding} \ast W +b,\\
    \mathop{\arg\max}\limits_{y_t}(\sum^n_{i=0}ABS(W_{L(i,y_t)})),
    \label{equ: trojan_neuron_selection}
\end{aligned}
\end{equation}

Once the dirty-label and corresponding target neurons are selected, the poison samples are generated as shown in Equation~\ref{equ: trojan_generation}, where $lr$ is the learning rate and $cost$ is the sum of the square error of target values and actual neuron values.

\begin{equation}
    x_p = x_b + (\mathcal{T} - lr\cdot\frac{\partial cost}{\partial \mathcal{T}}).
    \label{equ: trojan_generation}
\end{equation}

\subsection{Clean-Label Attacks}

\hspace{1.2em} \textbf{Sinusoidal Signal Backdoor (SIG).} SIG devises the sinusoidal signal as the backdoor trigger, expressed as:
\begin{equation}
    \mathcal{T}(i,j) = \Delta\mathop{\sin}(2{\pi}jf/m), 1\leqslant j \leqslant m, 1\leqslant i \leqslant l,
\end{equation}
where $f$ is the frequency of the sinusoidal signal, $\Delta$ is the strength, $m$ and $l$ are the numbers of columns and rows of an image, respectively. Then the backdoor trigger $\mathcal{T}$ is superimposed with benign images $x_b$ to form poisoned samples $x_p$, i.e.,  $x_p = x_b + \mathcal{T}$.

\textbf{Label Consistent Backdoor (LCBD).} LCBD places the trigger on images that are hard to classify so that the model relies more on the backdoor trigger to learn. The work proposes two approaches to make images hard-to-classify. The first generative adversarial network (GAN) based method crafts images $\tilde{x_p}$ by interpolating a given image $x_1$ to another $x_2$ through the latent space, where $x_1$ is the image from the poisoned class while $x_2$ is an arbitrary image from a different class. The process can be expressed as:
\begin{equation}
\begin{aligned}
    &\tilde{z_1} = \mathop{\arg\min}\limits_{z_1\in \mathds{R}^d}||x_1 - \mathcal{G}(z_1)||_2,\\
    &\tilde{z_2} = \mathop{\arg\min}\limits_{z_2\in \mathds{R}^d}||x_2 - \mathcal{G}(z_2)||_2,
\label{GAN-based-interpolation}
\end{aligned}
\end{equation}
\begin{equation}
    \tilde{x_p} = \mathcal{G}({\uptau}\tilde{z_1} + (1-{\uptau})\tilde{z_2}),
\label{GAN-based-interpolation-2}
\end{equation}
where $\tilde{z_1}$ and $\tilde{z_2}$ are the optimized latent space variables that produce close inputs to $x_1$ and $x_2$, respectively. $\mathcal{G}$ is the generator of a well-trained GAN model and $\uptau$ is the parameter to direct the transition from $x_1$ to $x_2$.

The second adversarial perturbation based method leverages the principle of adversarial examples (AE)~\cite{DBLP:journals/corr/GoodfellowSS14} and constructs hard-to-classify images $\tilde{x_p}$ as:
\begin{equation}
     \tilde{x_p} = \mathop{\arg\max} \mathcal{L}(x_b+\delta,y,\theta) \quad \text{s.t.} \quad ||\delta||_p \leqslant\epsilon,
\end{equation}
where $x_b$ represents benign images, $y$ represents the ground-truth labels, $\theta$ is the model parameters and $\delta$ is the adversarial perturbation constraint by the upper bound $\epsilon$ in $\ell_p$-norm. Lastly, poisoned samples are generated by placing a pre-defined trigger $\mathcal{T}$ on $\tilde{x_p}$:
\begin{equation}
    x_p = \mathcal{T} \odot \mathcal{M} + \tilde{x_p} \odot (1-\mathcal{M}),
\end{equation}

\textbf{Hidden Trigger Backdoor (HTBD).} HTBD is an even stealthier attack based on feature collision. The attack has two steps to generate poisoned data. As shown in Equations~(\ref{HTBD-step1}) and~(\ref{HTBD-step2}), it generates a base image (from the source class) attached with a pre-defined trigger and then optimize poisoned data $x_p$ (from the target class) by minimizing their $\ell_2$-norm distance with $\tilde{x_b}$ in the feature space.
\begin{equation}
    \tilde{x_b} = x_b \odot (1-\mathcal{M}) + \mathcal{T} \odot \mathcal{M},
\label{HTBD-step1}
\end{equation}
\begin{equation}
    x_p  = \mathop{\arg\max}\limits_{||x-x_t||_{\infty} < \epsilon} ||f(x) - f(\tilde{x_b})||^2_2 .
\label{HTBD-step2}
\end{equation}


\section{Details of Experimental Settings}
We introduce the details of experimental settings in this section. Following the open-sourced repositories of original papers~\cite{gu2017badnets,Trojannn,DBLP:journals/corr/abs-1712-05526,DBLP:conf/icip/BarniKT19,DBLP:journals/corr/abs-1912-02771,DBLP:conf/aaai/SahaSP20}, all experiments are implemented in Tensorflow~\cite{tensorflow2015-whitepaper} (LCBD\footnote{https://github.com/MadryLab/label-consistent-backdoor-code}) and PyTorch~\cite{paszke2019pytorch} (BadeNets\footnote{https://github.com/Kooscii/BadNets}, Blended, Trojan\footnote{https://github.com/PurduePAML/TrojanNN}, SIG\footnote{https://github.com/ebagdasa/backdoor\_federated\_learning} and HTBD\footnote{https://github.com/UMBCvision/Hidden-Trigger-Backdoor-Attacks}), and run on NVIDIA Tesla V100 GPUs. To reproduce the attacks and achieve the best attack performance, we follow original settings introduced in original papers. The statistics of datasets and models architecture for BadNets, Blended, Trojan, SIG, LCBD, and HTBD are summarized in Table~\ref{experiment-setting}. Note that all the datasets used in this paper, i.e., GTSRB\footnote{https://benchmark.ini.rub.de/gtsrb\_news.html}~\cite{stallkamp2011german}, CIFAR-10\footnote{https://www.cs.toronto.edu/~kriz/cifar.html}, and ImageNet-ILSVRC2012\footnote{https://www.image-net.org/challenges/LSVRC/2012/index.php}~\cite{deng2009imagenet}, are public.

The detailed training settings of each attack are described as follows.

\textbf{BadNets, Blended and Trojan.}
For the dirty-label attacks experiments, we use Stochastic Gradient Descent (SGD) optimizer with momentum of 0.9 and a weight decay of $5 \times 10^{-4}$ to train DNN models from scratch. The initial learning rate, batch size, and total training epochs are set to 0.1, 128, and 200, respectively.

\textbf{SIG.} We train DNN models from scratch using SGD optimizer with momentum of 0.9 in both pre-clean and post-clean phases. The initial learning rate is set to 0.01, with a weight decay of $5 \times 10^{-4}$. Models are trained for 100 epochs using a batch size of 32. 

\textbf{LCBD.} All models are trained from scratch using SGD with momentum of 0.9, a weight decay of $2 \times 10^{-4}$ and batch size of 50. We train models for 80000 steps in total with a learning rate that starts at 0.1 and schedule a learning rate decay that reduces the learning rate to 0.01 at 40000 steps and 0.001 at 60000 steps. 

\textbf{HTBD.} Different from SIG and LCBD, the HTBD trains models in a fine-tuning fashion according to the original paper~\cite{DBLP:conf/aaai/SahaSP20}. During the generation of the poisoned samples, we use the fc7 features of AlexNet (i.e., extract features from the 7th fully-connected layer of AlexNet) for feature collision. Poisoned data are generated with a learning rate of 0.01 in total 2 optimization epochs, and each epoch has 5000 iterations. During fine-tuning, models are trained using SGD with momentum of 0.9 and a learning rate of 0.001. The batch size and epoch are set to 256 and 30 for detection on the poisoned class and 1024 and 10 for the detection on the whole training dataset, respectively. 

\begin{table}[tbp]
\centering
\small

\begin{tabular}{cccc}
\toprule
\textbf{Attack} & \textbf{Dataset}  &\textbf{\begin{tabular}[c]{@{}c@{}}\#Training\\ Images\end{tabular}} & \textbf{\begin{tabular}[c]{@{}c@{}}DNN\\ Model\end{tabular}} \\ \midrule
SIG             & GTSRB                                                                               & 4,772                                                                 & ResNet                                                    \\

BadNet  &CIFAR-10       &50,000 & VGG16\\
Trojan  &CIFAR-10       &50,000 & VGG16\\
Blended (HK)  &CIFAR-10       &50,000 & VGG16\\
Blended (RP) &CIFAR-10      &50,000 & VGG16\\
LCBD            & CIFAR-10                                                                              & 50,000                                                               & ResNet                                                       \\
HTBD            & ImageNet                                                                          & \textgreater 1M                                               & AlexNet     \\
                                                  \bottomrule
\end{tabular}
\caption{Dataset and model architecture statistics}
\label{experiment-setting}
\end{table}




\begin{figure*}[tbp]
    \centering
    \resizebox{1\textwidth}{!}{
    \includegraphics{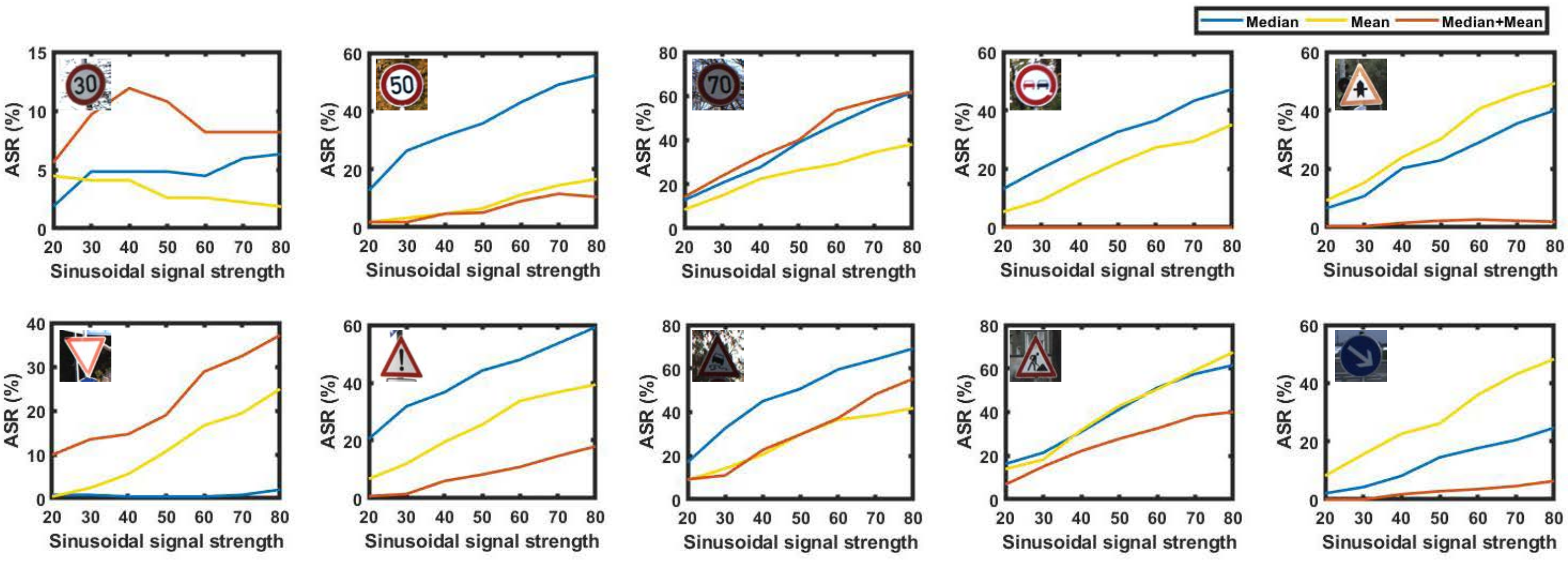}}
    \caption{ASR comparison of using single ``median'' baseline images, single ``mean'' baseline images and aggregation of median and mean (SIG). The aggregation of median and mean (red curves) achieves lower post-clean ASR in most cases, demonstrating a better backdoor mitigation capability over single ``median'' (blue curves) and single ``mean'' (yellow curves).}
    \label{fig:SIG_ASR_ablation}
\end{figure*}

\section{Ablation Study}
Our proposed methodology employs the natural property of DNNs and two denoising functions to detect the suspicious poisoned samples. Here, we examine the performance when applying only one of these two denoising functions to generate baseline images. Our experiment results show that different backdoor attacks have inconsistent sensitivity to various baseline images. In practical scenarios where the defender does not know the type of attack in advance, the safest strategy is to take baseline images generated by both denoising functions into consideration. Moreover, the ``median + mean'' version achieves better ASR reduction in general, although each single baseline version may have a higher detection rate in some cases.

We also evaluate the sensitivity of the removal threshold $\beta$. We find that only SIG is sensitive to $\beta$ while other attacks are not. In most cases, UltraClean can detect more than 80\% poisoned samples and reduce ASR significantly where $\beta$ is only at a value of 0.1. When $\beta$ approaches the value of 0.3, UltraClean detects nearly 100\% poisoned samples and almost completely removes the backdoor while maintaining a decent model accuracy. The experiment results provide insights into selecting $\beta$ in practice (i.e., a value of 0.3 would be good enough).

\subsection{Dirty-Label Attacks}
\hspace{1.2em}We present the ablation experiments results of dirty-label attacks in Table~\ref{abalation_dirty_label_attacks}. We find that BadNets is more sensitive to the type of baseline images. UltraClean fails to detect poisoned samples only based on baseline images generated by the mean denoising function, while successfully detecting most poisoned samples using baseline images generated by the median denoising function. On the other hand, UltraClean demonstrates consistently superior performance in detecting Blended and Trojan poisoned samples regardless of the type of baseline images. In general, the detection performance reaches the peak when we employ both baseline images. Meanwhile, all three attacks are not very sensitive to the change of the removal threshold. Although higher removal thresholds indeed improve the detection rate, a lower removal threshold can already effectively detect 83\%$\sim$97\% poisoned samples.

\begin{table}[tbp]
\centering
\small

\begin{tabular}{c|ccc}
\toprule
\textbf{\begin{tabular}[c]{@{}c@{}}Attack\\ Type\end{tabular}} & \textbf{\begin{tabular}[c]{@{}c@{}}BDR\\ (Median)\end{tabular}} & \textbf{\begin{tabular}[c]{@{}c@{}}BDR\\ (Mean)\end{tabular}} & \textbf{\begin{tabular}[c]{@{}c@{}}BDR\\ (Both)\end{tabular}} \\ \hline
\textbf{BadNets ($\boldsymbol\beta$ = 0.1)}                                        & 88.86\%                                                               & 0.04\%                                                                & 83.50\%                                                             \\
\textbf{BadNets ($\boldsymbol\beta$ = 0.2)}                                        & 94.52\%                                                               & 0.30\%                                                               & 92.74\%                                                             \\
\textbf{BadNets ($\boldsymbol\beta$ = 0.3)}                                        & 96.84\%                                                               & 1.48\%                                                               & 94.42\%                                                             \\\hline
\textbf{Trojan ($\boldsymbol\beta$ = 0.1)}                                         & 96.62\%                                                               & 79.84\%                                                             & 97.26\%                                                             \\
\textbf{Trojan ($\boldsymbol\beta$ = 0.2)}                                         & 99.10\%                                                               & 87.74\%                                                             & 99.32\%                                                             \\
\textbf{Trojan ($\boldsymbol\beta$ = 0.3)}                                         & 99.28\%                                                              & 90.58\%                                                             & 99.5\%                                                             \\\hline
\textbf{Blended (HK, $\boldsymbol\beta$ = 0.1)}                                   & 80.76\%                                                               & 82.20\%                                                             & 88.60\%                                                             \\
\textbf{Blended (HK, $\boldsymbol\beta$ = 0.2)}                                   & 93.18\%                                                               & 94.90\%                                                             & 95.84\%                                                            \\
\textbf{Blended (HK, $\boldsymbol\beta$ = 0.3)}                                   & 95.96\%                                                               & 96.60\%                                                             & 97.38\%                                                             \\\hline
\textbf{Blended (RP, $\boldsymbol\beta$ = 0.1)}                                   & 95.30\%                                                               & 82.68\%                                                             & 96.10\%                                                             \\
\textbf{Blended (RP, $\boldsymbol\beta$ = 0.2)}                                   & 99.18\%                                                               & 92.82\%                                                             & 99.46\%                                                            \\
\textbf{Blended (RP, $\boldsymbol\beta$ = 0.3)}                                   & 99.74\%                                                               & 95.52\%                                                             & 99.80\%                                                            \\ \bottomrule
\end{tabular}
\caption{Ablation study of denoising functions and $\beta$ on dirty-label attacks}
\label{abalation_dirty_label_attacks}
\end{table}

\subsection{Clean-Label Attacks}


\hspace{1.2em} \textbf{SIG.} The ablation study of baseline images generated by different denoising functions and $\beta$ for detection on the poisoned class and the whole dataset are summarized in Tables~\ref{ablation_SIG_1} and~\ref{ablation_SIG_2}, respectively. For the detection on the poisoned class, using the single ``mean'' baseline images achieves a slightly higher detection rate than using the single ``median'' function and the aggregation of mean and median for most classes. Results of the detection on the whole dataset reveal the same trend. We then compare the post-clean ASR and present the results in Figure~\ref{fig:SIG_ASR_ablation}. Although using the single ``mean'' baselines perform better in detection rate, there is still a considerable performance gap to applying both baseline images in most cases. In general, the aggregation of median and mean denoising shows the best performance in degrading the post-clean ASR. We believe this is because the aggregation leads to a larger noise difference, which makes poisoned samples easier to detect. The ablation study of $\beta$ shows that UltraClean's performance against SIG is sensitive to the adjustment of the removal threshold. A larger $\beta$ can detect more poisoned samples and better reduce ASR.

\begin{table}[tbp]
\centering
\small

\begin{tabular}{c|ccc}
\toprule
\textbf{Class ID} & \textbf{\begin{tabular}[c]{@{}c@{}}BDR\\ (Median)\end{tabular}} & \textbf{\begin{tabular}[c]{@{}c@{}}BDR\\ (Mean)\end{tabular}} & \textbf{\begin{tabular}[c]{@{}c@{}}BDR\\ (Both)\end{tabular}} \\ \hline
1                 & 52.17\%                                                                           & 63.48\%                                                                             & 52.17\%                                                                          \\
2                 & 56.41\%                                                                            & 65.38\%                                                                              & 56.41\%                                                                           \\
3                 & 47.54\%                                                                            & 65.57\%                                                                              & 50.82\%                                                                           \\
4                 & 67.22\%                                                                            & 80.00\%                                                                              & 69.01\%                                                                           \\
5                 & 47.29\%                                                                           & 57.43\%                                                                            & 46.62\%                                                                         \\
6                 & 48.71\%                                                                           & 70.77\%                                                                            & 52.82\%                                                                         \\
7                 & 49.61\%                                                                          & 70.08\%                                                                             & 62.99\%                                                                          \\
8                 & 51.92\%                                                                            & 51.92\%                                                                              & 51.92\%                                                                           \\
9                 & 46.23\%                                                                           & 62.81\%                                                                            & 48.74\%                                                                          \\
10                & 85.42\%                                                                            & 77.08\%                                                                              & 91.67\%                                                                          \\ \bottomrule
\end{tabular}
\caption{Ablation study of denoising functions for detection on the poisoned class (SIG)}
\label{ablation_SIG_1}
\end{table}

\begin{table}[tbp]
\centering
\small

\begin{tabular}{c|ccc}
\toprule
\textbf{\begin{tabular}[c]{@{}c@{}}$\boldsymbol\beta$\end{tabular}} & \textbf{\begin{tabular}[c]{@{}c@{}}BDR\\ (Median)\end{tabular}} & \textbf{\begin{tabular}[c]{@{}c@{}}BDR\\ (Mean)\end{tabular}} & \textbf{\begin{tabular}[c]{@{}c@{}}BDR\\ (Both)\end{tabular}} \\ \hline
0.00                                                                 & 0.00\%                                                                            & 0.00\%                                                                              & 0.00\%                                                                           \\
0.05                                                                 & 3.38\%                                                                            & 6.76\%                                                                             & 3.38\%                                                                           \\
0.10                                                                 & 9.46\%                                                                           & 11.49\%                                                                             & 9.46\%                                                                          \\
0.15                                                                 & 14.86\%                                                                          & 19.59\%                                                                             & 15.54\%                                                                          \\
0.20                                                                 & 25.67\%                                                                           & 29.73\%                                                                             & 25.00\%
\\
0.25                                                                 & 29.05\%                                                                          & 36.49\%                                                                             & 29.73\%                                                                          \\
0.30                                                                 & 41.89\%                                                                           & 42.57\%                                                                             & 39.86\%                                                                          \\ \bottomrule
\end{tabular}
\caption{Ablation study of denoising functions and $\beta$ for detection on the whole training dataset (SIG)}
\label{ablation_SIG_2}
\end{table}

\begin{table}[tbp]
\centering
\small

\begin{tabular}{c|ccc}
\toprule
\textbf{\begin{tabular}[c]{@{}c@{}}Attack type\end{tabular}} & \textbf{\begin{tabular}[c]{@{}c@{}} BDR\\(Median)\end{tabular}} & \textbf{\begin{tabular}[c]{@{}c@{}}BDR\\(Mean)\end{tabular}} & \textbf{\begin{tabular}[c]{@{}c@{}}BDR\\(Both)\end{tabular}} \\ \hline
\textbf{GAN ($\uptau$ = 0.0)}                                            & 75.60\%                                                                         & 51.20\%                                                                         & 72.70\%                                                                    \\
\textbf{GAN ($\uptau$ = 0.1)}                                            & 73.25\%                                                                        & 41.80\%                                                                          & 70.95\%                                                                    \\
\textbf{GAN ($\uptau$ = 0.2)}                                            & 80.15\%                                                                        & 40.20\%                                                                          & 78.00\%                                                                    \\
\textbf{GAN ($\uptau$ = 0.3)}                                            & 83.15\%                                                                        & 41.55\%                                                                          & 79.75\%                                                                    \\ \hline
\textbf{AE ($\ell_2$, $\epsilon$ = 300)}                                          & 77.65\%                                                                        & 36.85\%                                                                          & 75.55\%                                                                    \\
\textbf{AE ($\ell_2$, $\epsilon$ = 600)}                                          & 89.45\%                                                                        & 45.95\%                                                                          & 88.65\%                                                                    \\
\textbf{AE ($\ell_2$, $\epsilon$ = 1200)}                                         & 98.80\%                                                                        & 78.75\%                                                                          & 98.55\%                                                                   \\ \hline
\textbf{AE ($\ell_\infty$, $\epsilon$ = 8)}                                            & 77.55\%                                                                        & 42.05\%                                                                          & 75.80\%                                                                   \\
\textbf{AE ($\ell_\infty$, $\epsilon$ = 16)}                                           & 89.40\%                                                                         & 40.95\%                                                                          & 88.55\%                                                                    \\
\textbf{AE ($\ell_\infty$, $\epsilon$ = 32)}                                           & 97.40\%                                                                        & 44.75\%                                                                          & 97.15\%                                                                    \\ \bottomrule
\end{tabular}
\caption{Ablation study of denoising functions for detection on the poisoned class (LCBD)}
\label{ablation_LCBD_1}
\end{table}

\begin{table}[tbp]
\centering
\small

\begin{tabular}{c|ccc}
\toprule
\textbf{\begin{tabular}[c]{@{}c@{}}$\boldsymbol\beta$\end{tabular}} & \textbf{\begin{tabular}[c]{@{}c@{}}BDR\\ (Median)\end{tabular}} & \textbf{\begin{tabular}[c]{@{}c@{}}BDR\\ (Mean)\end{tabular}} & \textbf{\begin{tabular}[c]{@{}c@{}}BDR\\ (Both)\end{tabular}} \\ \hline
0.00                                                                 & 0.00\%                                                                & 0.00\%                                                                 & 0.00\%                                                                 \\
0.05                                                                 & 77.35\%                                                             & 0.00\%                                                                & 59.50\%                                                              \\
0.10                                                                 & 89.40\%                                                              & 0.30\%                                                                 & 88.80\%                                                              \\
0.15                                                                 & 93.25\%                                                              & 0.30\%                                                                 & 93.00\%                                                              \\
0.20                                                                 & 94.90\%                                                              & 0.60\%                                                                & 94.65\%                                                              \\
0.25                                                                 & 96.05\%                                                              & 1.15\%                                                                & 96.35\%                                                             \\
0.30                                                                 & 97.65\%                                                              & 2.50\%                                                                & 97.40\%                                                              \\ \bottomrule
\end{tabular}
\caption{Ablation study of denoising functions and $\beta$ for detection on the whole training dataset (LCBD)}
\label{ablation_LCBD_2}
\end{table}

\textbf{LCBD.} Tables~\ref{ablation_LCBD_1} and~\ref{ablation_LCBD_2} present the detection performance against LCBD on the poisoned class and the whole dataset, respectively. It can be seen that the single ``median'' denoised baseline images outperform the single ``mean'' denoised baseline images by a large margin in detecting poisoned samples formed by LCBD, which is different from the SIG attacks. On the other hand, the aggregation of median and mean denoising still has a comparable detection rate to the single ``median'' denoising. This trend is even prominent for the detection on the whole dataset where employing the single ``mean'' denoising images can barely detect any poisoned sample. Meanwhile, we found the post-clean ASR of employing the single ``median'' and the aggregation version are close, since they detect almost the same poisoned samples. In contrast to SIG, defending against LCBD is not too sensitive to the selection of $\beta$. As can be seen in Table~\ref{ablation_LCBD_2}, UltraClean still detects 88.80\% poisoned samples when $\beta$ is small.

\begin{figure*}[htbp]
    \centering
    \resizebox{1\textwidth}{!}{
    \includegraphics{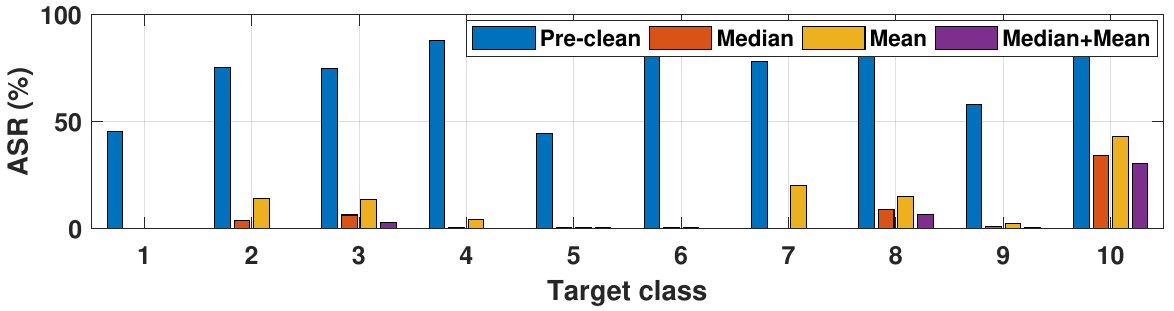}}
    \caption{ASR comparison of using the single ``median'' baseline images, single ``mean'' baseline images and aggregation of median and mean (HTBD).}
    \label{fig:HTBD_ASR_ablation}
\end{figure*}

\begin{table}[htbp]
\centering
\small

\begin{tabular}{c|c|ccc}
\toprule
\multicolumn{2}{c|}{\textbf{Target Class}}          & \textbf{\begin{tabular}[c]{@{}c@{}}BDR\\ (Median)\end{tabular}} & \textbf{\begin{tabular}[c]{@{}c@{}}BDR\\ (Mean)\end{tabular}} & \textbf{\begin{tabular}[c]{@{}c@{}}BDR\\ (Both)\end{tabular}} \\ \hline
1& Terrier       & 92.00\%                                                                  & 86.00\%                                                                & 97.00\%                                                               \\
2& Bee                      & 84.00\%                                                                  & 86.00\%                                                                & 86.00\%                                                                \\
3& Plunger                  & 84.00\%                                                                  & 79.00\%                                                                & 86.00\%                                                                \\
4& Partridge                & 85.00\%                                                                  & 78.00\%                                                             & 87.00\%                                                                \\
5& Ipod                     & 91.00\%                                                                 & 92.00\%                                                                & 96.00\%                                                                \\
6& Deerhound       & 100.00\%                                                                 & 72.00\%                                                                & 96.00\%                                                                \\
7& Cockatoo & 99.00\%                                                                  & 91.00\%                                                                & 99.00\%                                                                \\
8& Toyshop                  & 71.00\%                                                                  & 70.00\%                                                                & 72.00\%                                                                \\
9& Tiger beetle             & 100.00\%                                                                 & 94.00\%                                                                & 100.00\%                                                               \\
10& Goblet                   & 79.00\%                                                                  & 69.00\%                                                                & 86.00\%                                                                \\ \bottomrule
\end{tabular}
\caption{Ablation study of denoising functions for detection on the poisoned class (HTBD)}
\label{ablation_HTBD_1}
\end{table}

\textbf{HTBD.} Unlike SIG and LCBD, for the detection on the poisoned class, HTBD attacks are sensitive to the single ``median'' and the single ``mean'' denoised baseline images. However, the aggregation version consistently achieves the highest detection rate for all target classes, as shown in Table~\ref{ablation_HTBD_1}. We can observe from Figure~\ref{fig:HTBD_ASR_ablation} that the performance of the post-clean ASR is similar to the detection rate, where the aggregation version always performs the best. For the detection on the whole dataset, applying both the aggregation version and the single ``median'' denoised baseline images detect considerable poisoned samples while applying the single ``mean'' denoising only detects a few of them. UltraClean is resistant to the selection of $\beta$ in defending against HTBD, only a removal threshold of 5\% is capable of entirely mitigating HTBD from models.

\begin{table}[tbp]
\centering
\small

\begin{tabular}{c|ccc}
\toprule
\textbf{\begin{tabular}[c]{@{}c@{}}$\boldsymbol\beta$\end{tabular}} & \textbf{\begin{tabular}[c]{@{}c@{}}BDR\\ (Median)\end{tabular}} & \textbf{\begin{tabular}[c]{@{}c@{}}BDR\\ (Mean)\end{tabular}} & \textbf{\begin{tabular}[c]{@{}c@{}}BDR\\ (Both)\end{tabular}} \\ \hline
0.00                                                                 & 0.00\%                                                                 & 0.00\%                                                                 & 0.00\%                                                                 \\
0.01                                                                 & 65.50\%                                                              & 12.00\%                                                                 & 57.25\%                                                              \\
0.02                                                                 & 79.25\%                                                              & 19.75\%                                                                & 68.75\%                                                              \\
0.03                                                                 & 87.00\%                                                              & 25.25\%                                                                 & 78.00\%                                                              \\
0.04                                                                 & 91.00\%                                                              & 29.50\%                                                                & 83.50\%                                                             \\
0.05                                                                 & 93.25\%                                                              & 33.25\%                                                                & 88.25\%                                                         \\ \bottomrule
\end{tabular}
\caption{Ablation study of denoising functions and $\beta$ for detection on the whole training dataset (HTBD)}
\label{ablation_HTBD_2}
\end{table}

In sum, all six dirty-label and clean-label attacks reveal inconsistent sensitivity to different denoised baseline images. Only the aggregation of both functions demonstrates effectiveness against all the attacks. Overall, the differences of exploiting each single denoised baseline image in detecting different backdoor attacks further validate the necessity of integrating both denoising functions in the proposed methodology. Similarly, UltraClean's sensitivity to removal threshold is different against different attacks. However, a value of 0.3 is good enough to handle all attacks.


\begin{table}[htbp]
\centering
\small

\begin{tabular}{cc|ccc}
\toprule
\multicolumn{2}{c|}{\textbf{\begin{tabular}[c]{@{}c@{}}Attack\\ Type\end{tabular}}}                  & \textbf{\begin{tabular}[c]{@{}c@{}}Acc.\\ (UC)\end{tabular}} & \textbf{\begin{tabular}[c]{@{}c@{}}BDR\\ (UC)\end{tabular}} & \textbf{\begin{tabular}[c]{@{}c@{}}ASR\\ (UC)\end{tabular}} \\ \hline
\multicolumn{1}{c|}{\multirow{3}{*}{\textbf{BadNets}}}                                                   & \textbf{$\mathcal{T}_s$=1x1}               &      83.49\%                                                        &   92.34\%                                                                  &      0.59\%                                                       \\
\multicolumn{1}{c|}{}                                                                                    & \textbf{$\mathcal{T}_s$=2x2}             &         84.34\%                                                     &  100.00\%                                                                   &     0.82\%                                                        \\
\multicolumn{1}{c|}{}                                                                                    & \textbf{$\mathcal{T}_s$=3x3}            & 83.91\%                                                             &94.42\%                                                                     &    0.83\%                                                         \\ \hline
\multicolumn{1}{c|}{\multirow{3}{*}{\textbf{Trojan}}}                                                    & \textbf{$\mathcal{T}_t$=0.1}           &       84.83\%                                                       &  32.34\%                                                                   &   0.36\%                                                          \\
\multicolumn{1}{c|}{}                                                                                    & \textbf{$\mathcal{T}_t$=0.3}             &          84.83\%                                                    &   98.66\%                                                                 &   0.40\%                                                          \\
\multicolumn{1}{c|}{}                                                                                    & \textbf{$\mathcal{T}_t$=0.5}            &   84.73\%                                                           & 99.50\%                                                                    &  1.61\%                                                           \\ \hline
\multicolumn{1}{c|}{\multirow{3}{*}{\textbf{\begin{tabular}[c]{@{}c@{}}Blended\\ \\ (HK)\end{tabular}}}} & \textbf{$\alpha$=0.1}            &     84.34\%                                                         &  96.00\%                                                                  &  1.30\%                                                           \\
\multicolumn{1}{c|}{}                                                                                    & \textbf{$\alpha$=0.2}             &   85.08\%                                                           &  97.38\%                                                                   &   3.06\%                                                          \\
\multicolumn{1}{c|}{}                                                                                    & \textbf{$\alpha$=0.3}            &      85.00\%                                                        &   98.44\%                                                                  &   10.23\%                                                          \\ \hline
\multicolumn{1}{c|}{\multirow{3}{*}{\textbf{\begin{tabular}[c]{@{}c@{}}Blended\\ \\ (RP)\end{tabular}}}} & \textbf{$\alpha$=0.1}         &    84.23\%                                                          &   99.32\%                                                                  &  0.70\%                                                           \\
\multicolumn{1}{c|}{}                                                                                    & \textbf{$\alpha$=0.2}           &    84.23\%                                                          &  99.80\%                                                                  &    0.15\%                                                         \\
\multicolumn{1}{c|}{}                                                                                    & \textbf{$\alpha$=0.3}           &    83.92\%                                                          &  99.78\%                                                                   &     0.11\%                                                        \\ \bottomrule
\end{tabular}
\caption{UltraClean against adaptive attacks (dirty-label)}
\label{adaptive_dirty_label}
\end{table}

\section{Robustness against Adaptive Attacks}

Attackers may perform adaptive attacks by adjusting the poisoning ratio, changing trigger size, varying blended or transparency ratio, or altering adversarial perturbation to evade defenses. We extensively evaluate UltraClean's robustness to multiple adaptive attacks and find none of the attacks successfully bypass the detection of UltraClean. UltraClean consistently detects most poisoned samples and mitigates ASR by a large margin against both dirty-label and clean-label adaptive attacks. Detailed results are presented as follows.

\subsection{Adaptive Attacks of Dirty-Label Attacks}
\hspace{1.2em} We first evaluate UltraClean's performance against adaptive attacks of dirty-label attacks. In order to bypass the defense of UltraClean, attackers can reduce backdoor trigger size or trigger blended/transparency ratio to craft harder-to-detect poisoned samples. Note that attackers tend not to reduce the poisoning ratio for dirty-label attacks since it is usually below 10\% to evade label sanitization defense~\cite{DBLP:journals/corr/abs-1803-00992}. A lower poisoning ratio may render a failure of attacks. We conduct the adaptive attacks by adjusting the trigger size ($\mathcal{T}_s$) of BadNets from $1\times 1$ to $3\times 3$, the trigger transparency ($\mathcal{T}_t$) of Trojan from 0.1 to 0.5 and blended ratio ($\alpha$) of Blended from 0.1 to 0.3. The results are summarized in Table~\ref{abalation_dirty_label_attacks}. We can see that while most attacks can achieve high ASR even with harder-to-detect triggers, UltraClean demonstrates exceptional robustness to various adaptive attacks. UltraClean identifies at least 92\% poisoned samples for almost all adaptive attacks and successfully mitigates backdoors in all models. The only exception is the Trojan attack with $\mathcal{T}_t = 0.1$, where the attack achieves 9.43\% ASR even without a defense. Although UltraClean only detects 32\% of the poisoned data, it still completely remove the backdoor. \textbf{The experiment results validates UltraClean's robustness against dirty-label adaptive attacks.}

\begin{table}[htbp]
\centering
\small

\begin{tabular}{c|ccccc}
\toprule
\textbf{\begin{tabular}[c]{@{}c@{}}Poisoning\\ Ratio\end{tabular}}   & \textbf{\begin{tabular}[c]{@{}c@{}}Acc.\\ (PC)\end{tabular}} & \textbf{\begin{tabular}[c]{@{}c@{}}ASR\\ (PC)\end{tabular}} & \textbf{\begin{tabular}[c]{@{}c@{}}BDR\\ (UC)\end{tabular}} & \textbf{\begin{tabular}[c]{@{}c@{}}Acc.\\ (UC)\end{tabular}} & \textbf{\begin{tabular}[c]{@{}c@{}}ASR\\ (UC)\end{tabular}} \\ \hline
\textbf{0.1}                                                         & 88.48\%                                                             & 99.01\%                                                            & 95.80\%                                                               & 88.27\%                                                        & 1.69\%                                                        \\
\textbf{0.2}                                                         & 87.98\%                                                             & 99.76\%                                                            & 97.30\%                                                             & 87.89\%                                                        & 1.49\%                                                        \\
\textbf{0.3}                                                         & 88.02\%                                                             & 99.98\%                                                            & 97.06\%                                                             & 87.81\%                                                        & 1.38\%                                                        \\ 
\textbf{0.4}                                                         & 87.73\%                                                             & 99.98\%                                                            & 98.55\%                                                            & 87.26\%                                                        & 1.10\%                                                        \\\hline
\textbf{\begin{tabular}[c]{@{}c@{}}Trigger\\ Amplitude\end{tabular}} & \textbf{\begin{tabular}[c]{@{}c@{}}Acc.\\ (PC)\end{tabular}} & \textbf{\begin{tabular}[c]{@{}c@{}}ASR\\ (PC)\end{tabular}} & \textbf{\begin{tabular}[c]{@{}c@{}}BDR\\ (UC)\end{tabular}} & \textbf{\begin{tabular}[c]{@{}c@{}}Acc.\\ (UC)\end{tabular}} & \textbf{\begin{tabular}[c]{@{}c@{}}ASR\\ (UC)\end{tabular}} \\ \hline
\textbf{16}                                                          & 87.44\%                                                             & 1.12\%                                                             & 90.35\%                                                             & 86.45\%                                                        & 1.04\%                                                        \\
\textbf{32}                                                          & 87.39\%                                                             & 99.90\%                                                            & 98.70\%                                                             & 87.32\%                                                        & 0.88\%                                                        \\
\textbf{64}                                                          & 87.73\%                                                             & 99.98\%                                                            & 98.55\%                                                             & 87.26\%                                                        & 1.10\%                                                        \\\bottomrule
\end{tabular}
\caption{UltraClean against adaptive attacks (LCBD)}
\label{LCBD_different_setting}
\end{table}

\begin{table}[htbp]
\centering
\small

\begin{tabular}{c|c|cccc}
\toprule
\multicolumn{2}{c|}{\textbf{Target Class}}   &  \textbf{\begin{tabular}[c]{@{}c@{}}BDR\\ ($\epsilon$=8) \\ ($\mathcal{T}_s$=30)\end{tabular}} & \textbf{\begin{tabular}[c]{@{}c@{}}BDR\\ ($\epsilon$=32)\\($\mathcal{T}_s$=30)\end{tabular}} & \textbf{\begin{tabular}[c]{@{}c@{}}BDR\\ ($\epsilon$=16)\\($\mathcal{T}_s$=15)\end{tabular}} & \textbf{\begin{tabular}[c]{@{}c@{}}BDR\\ ($\epsilon$=16)\\($\mathcal{T}_s$=60)\end{tabular}} \\ \hline
1& Terrier                                                                                    & 100.00\%                                                                             & 100.00\%                                                                               & 99.50\%                                                                               &   99.75\%                                                                                    \\
2& Bee                                                                                                    & 95.25\%                                                                              & 96.50\%                                                                               & 93.75\%                                                                               &    93.50\%                                                                                   \\
3& Plunger                                                                                                & 94.00\%                                                                              & 94.75\%                                                                               & 94.25\%                                                                               &   94.00\%                                                                                   \\
4& Partridge                                                                                              & 97.50\%                                                                              & 96.25\%                                                                               & 96.00\%                                                                               &    97.00\%                                                                                   \\
5& Ipod                                                                                                   & 99.25\%                                                                              & 99.00\%                                                                              & 99.00\%                                                                               &    99.50\%                                                                                   \\
6& Deerhound                                                                                     & 99.75\%                                                                             & 98.75\%                                                                               & 100.00\%                                                                              &   97.25\%                                                                                     \\
7& Cockatoo                                                                               & 99.75\%                                                                              & 99.25\%                                                                               & 96.00\%                                                                               &  99.00\%                                                                                     \\
8& Toyshop                                                                                               & 89.50\%                                                                              & 88.25\%                                                                               & 92.75\%                                                                              &   88.25\%                                                                                    \\
9& Tiger beetle                                                                                           & 98.00\%                                                                              & 97.75\%                                                                               & 99.00\%                                                                               &    96.25\%                                                                                   \\
10& Goblet                                                                                                 & 94.75\%                                                                              & 95.00\%                                                                               & 92.75\%                                                                               &  93.75\%                                                                                     \\ \bottomrule
\end{tabular}
\caption{Detection performance of UltraClean against adaptive attacks (HTBD)}
\label{HTBD_different_settings}
\end{table}

\subsection{Adaptive Attacks of Clean-Label Attacks}
\hspace{1.2em} \textbf{LCBD.} In the main manuscript, we extensively evaluated UltraClean against LCBD with GAN-based and AE-based approaches. UltraClean has shown superior performance in detecting and mitigating the LCBD backdoor. Here, we further study the effectiveness of UltraClean against adaptive attacks. Attackers may change poisoning ratio and trigger amplitude to make LCBD more stealthy. In the previous experiments, we injected 40\% poisoned sample into the target class. Although the poisoning ratio (i.e., the fraction of poisoned data injected into the training dataset) over the entire dataset is small, 40\% is a relatively high poisoning ratio over the poisoned class. The original LCBD paper~\cite{DBLP:journals/corr/abs-1912-02771} shows that with lower poisoning ratios (lower poisoning ratios usually render higher stealthiness), it can still achieve a high attack success rate. Thus, we investigate if UltraClean is still effective with lower poisoning ratios. We perform the experiment with the setting ``AE ($\ell_2$, $\epsilon$ = 1200)'' since it achieves the best attack result. As shown in Table~\ref{LCBD_different_setting}, even with 10\% poisoned data in the target class (only 1\% poisoning ratio over the entire training dataset), LCBD still achieves $\sim$99\% ASR. Even under this circumstance, UltraClean still demonstrates a stable detection performance, reaching at least a 95\% detection rate and reducing the ASR by over 98\%. Another critical factor to make the attack more stealthy is the trigger amplitude. Smaller amplitude means less trigger visibility. In the previous experiments, we set the amplitude to 64 to secure the attack success rate. Here we reduce the amplitude to 16 and 32 to examine if UltraClean can still detect poisoned samples. The experiment results are summarized in the last two rows in Table~\ref{LCBD_different_setting}.  We find that UltraClean still detects at least 90\% poisoned samples, even with an extremely low trigger amplitude. Note that with a 16 trigger amplitude, the attack has failed to inject an effective backdoor since the pre-clean ASR is only 1.12\%. \textbf{The results further demonstrate the effectiveness of UltraClean against different attack settings, indicating its robustness in protecting neural networks against various adaptive attacks.}

\begin{table*}[htbp]
\centering
\setlength{\tabcolsep}{8pt} 
\resizebox{\linewidth}{!}{
\begin{tabular}{c|cccccccccc}
\toprule
\multirow{2}{*}{\textbf{\begin{tabular}[c]{@{}c@{}}$\epsilon$=16\\ $\mathcal{T}_s$=30\end{tabular}}} & \multicolumn{10}{c}{\textbf{Class ID}}                                                                                                                                         \\ \cline{2-11} 
                                                                               & \textbf{1}      & \textbf{2}      & \textbf{3}     & \textbf{4}      & \textbf{5}      & \textbf{6}      & \textbf{7}     & \textbf{8}     & \textbf{9}      & \textbf{10}     \\ \hline
\textbf{\begin{tabular}[c]{@{}c@{}}Acc.(\%)\\ (Pre)\end{tabular}}       & 96.00           & 97.00           & 95.00          & 97.00           & 95.00           & 95.00           & 96.00          & 95.00          & 98.00           & 95.00           \\ \cline{1-1}
\textbf{\begin{tabular}[c]{@{}c@{}}Acc.(\%)\\ (UC)\end{tabular}}            & \textbf{100.00} & \textbf{99.00}  & \textbf{97.00} & \textbf{100.00} & \textbf{100.00} & \textbf{100.00} & \textbf{98.00} & \textbf{98.00} & \textbf{100.00} & \textbf{99.00}  \\ \cline{1-1}
\textbf{\begin{tabular}[c]{@{}c@{}}ASR(\%)\\ (Pre)\end{tabular}}        & 45.25           & 75.00           & 74.5           & 87.75           & 44.50           & 83.75           & 78.00          & 80.50          & 58.00           & 89.75           \\ \cline{1-1}
\textbf{\begin{tabular}[c]{@{}c@{}}ASR(\%)\\ (UC)\end{tabular}}             & \textbf{0.00}   & \textbf{0.00}   & \textbf{4.75}  & \textbf{0.00}   & \textbf{0.00}   & \textbf{0.25}   & \textbf{0.00}  & \textbf{4.50}  & \textbf{0.00}   & \textbf{27.50}  \\ \hline
\multirow{2}{*}{\textbf{\begin{tabular}[c]{@{}c@{}}$\epsilon$=8\\ $\mathcal{T}_s$=30\end{tabular}}}  & \multicolumn{10}{c}{\textbf{Class ID}}                                                                                                                                         \\ \cline{2-11} 
                                                                               & \textbf{1}      & \textbf{2}      & \textbf{3}     & \textbf{4}      & \textbf{5}      & \textbf{6}      & \textbf{7}     & \textbf{8}     & \textbf{9}      & \textbf{10}     \\ \hline
\textbf{\begin{tabular}[c]{@{}c@{}}Acc.(\%)\\ (Pre)\end{tabular}}       & 98.00           & 99.00           & 95.00          & 98.00           & 96.00           & 90.00           & 96.00          & 96.00          & 99.00           & 98.00           \\ \cline{1-1}
\textbf{\begin{tabular}[c]{@{}c@{}}Acc.(\%)\\ (UC)\end{tabular}}      & \textbf{100.00} & \textbf{100.00} & \textbf{97.00} & \textbf{100.00} & \textbf{100.00} & \textbf{99.00}  & \textbf{99.00} & \textbf{96.00} & \textbf{100.00} & \textbf{99.00}  \\ \cline{1-1}
\textbf{\begin{tabular}[c]{@{}c@{}}ASR(\%)\\ (Pre)\end{tabular}}        & 33.75           & 78.75           & 78.25          & 85.25           & 80.00           & 41.75           & 32.00          & 79.50          & 87.00           & 89.75           \\ \cline{1-1}
\textbf{\begin{tabular}[c]{@{}c@{}}ASR(\%)\\ (UC)\end{tabular}}       & \textbf{0.00}   & \textbf{0.00}   & \textbf{2.75}  & \textbf{0.00}   & \textbf{0.50}   & \textbf{0.00}   & \textbf{0.00}  & \textbf{6.50}  & \textbf{0.50}   & \textbf{3.25} \\ \hline
\multirow{2}{*}{\textbf{\begin{tabular}[c]{@{}c@{}}$\epsilon$=32\\ $\mathcal{T}_s$=30\end{tabular}}} & \multicolumn{10}{c}{\textbf{Class ID}}                                                                                                                                         \\ \cline{2-11} 
                                                                               & \textbf{1}      & \textbf{2}      & \textbf{3}     & \textbf{4}      & \textbf{5}      & \textbf{6}      & \textbf{7}     & \textbf{8}     & \textbf{9}      & \textbf{10}     \\ \hline
\textbf{\begin{tabular}[c]{@{}c@{}}Acc.(\%)\\ (Pre)\end{tabular}}       & 97.00           & 99.00           & 96.00          & 99.00           & 93.00           & 91.00           & 98.00          & 95.00          & 100.00          & 95.00           \\ \cline{1-1}
\textbf{\begin{tabular}[c]{@{}c@{}}Acc.(\%)\\ (UC)\end{tabular}}      & \textbf{100.00} & \textbf{100.00} & \textbf{97.00} & \textbf{100.00} & \textbf{100.00} & \textbf{99.00}  & \textbf{99.00} & \textbf{96.00} & \textbf{100.00} & \textbf{99.00}  \\ \cline{1-1}
\textbf{\begin{tabular}[c]{@{}c@{}}ASR(\%)\\ (Pre)\end{tabular}}        & 29.00           & 80.75           & 81.25          & 88.00           & 77.50           & 72.25           & 35.50          & 83.25          & 85.25           & 87.00           \\ \cline{1-1}
\textbf{\begin{tabular}[c]{@{}c@{}}ASR(\%)\\ (UC)\end{tabular}}       & \textbf{0.00}   & \textbf{0.00}   & \textbf{3.50}  & \textbf{0.00}   & \textbf{0.50}   & \textbf{0.00}   & \textbf{0.00}  & \textbf{7.75}  & \textbf{0.00}   & \textbf{1.50}   \\ \hline
\multirow{2}{*}{\textbf{\begin{tabular}[c]{@{}c@{}}$\epsilon$=16\\ $\mathcal{T}_s$=15\end{tabular}}} & \multicolumn{10}{c}{\textbf{Class ID}}                                                                                                                                         \\ \cline{2-11} 
                                                                               & \textbf{1}      & \textbf{2}      & \textbf{3}     & \textbf{4}      & \textbf{5}      & \textbf{6}      & \textbf{7}     & \textbf{8}     & \textbf{9}      & \textbf{10}     \\ \hline
\textbf{\begin{tabular}[c]{@{}c@{}}Acc.(\%)\\ (Pre)\end{tabular}}            & 97.00           & 98.00           & 96.00          & 99.00           & 96.00           & 92.00           & 97.00          & 97.00          & 100.00          & 96.00           \\ \cline{1-1}
\textbf{\begin{tabular}[c]{@{}c@{}}Acc.(\%)\\ (UC)\end{tabular}}           & \textbf{100.00} & \textbf{100.00} & \textbf{97.00} & \textbf{100.00} & \textbf{99.00}  & \textbf{100.00} & \textbf{98.00} & \textbf{96.00} & \textbf{100.00} & \textbf{99.00}  \\ \cline{1-1}
\textbf{\begin{tabular}[c]{@{}c@{}}ASR(\%)\\ (Pre)\end{tabular}}             & 40.25           & 66.25           & 80.75          & 89.00           & 75.00           & 30.75           & 86.00          & 77.75          & 84.75           & 97.00           \\ \cline{1-1}
\textbf{\begin{tabular}[c]{@{}c@{}}ASR(\%)\\ (UC)\end{tabular}}            & \textbf{0.00}   & \textbf{8.50}   & \textbf{4.50}  & \textbf{0.00}   & \textbf{0.50}   & \textbf{0.00}   & \textbf{0.00}  & \textbf{8.75}  & \textbf{0.25}   & \textbf{6.00}   \\ \hline
\multirow{2}{*}{\textbf{\begin{tabular}[c]{@{}c@{}}$\epsilon$=16\\ $\mathcal{T}_s$=60\end{tabular}}} & \multicolumn{10}{c}{\textbf{Class ID}}                                                                                                                                         \\ \cline{2-11} 
                                                                               & \textbf{1}      & \textbf{2}      & \textbf{3}     & \textbf{4}      & \textbf{5}      & \textbf{6}      & \textbf{7}     & \textbf{8}     & \textbf{9}      & \textbf{10}     \\ \hline
\textbf{\begin{tabular}[c]{@{}c@{}}Acc.(\%)\\ (Pre)\end{tabular}}            & 96.00           & 99.00           & 96.00          & 98.00           & 97.00           & 93.00           & 97.00          & 96.00          & 99.00           & 96.00           \\ \cline{1-1}
\textbf{\begin{tabular}[c]{@{}c@{}}Acc.(\%)\\ (UC)\end{tabular}}           & \textbf{100.00} & \textbf{100.00} & \textbf{97.00} & \textbf{100.00} & \textbf{100.00} & \textbf{99.00}  & \textbf{99.00} & \textbf{96.00} & \textbf{100.00} & \textbf{99.00}  \\ \cline{1-1}
\textbf{\begin{tabular}[c]{@{}c@{}}ASR(\%)\\ (Pre)\end{tabular}}             & 47.50           & 78.25           & 77.25          & 63.75           & 53.75           & 83.00           & 77.75          & 83.75          & 85.75           & 91.25           \\ \cline{1-1}
\textbf{\begin{tabular}[c]{@{}c@{}}ASR(\%)\\ (UC)\end{tabular}}            & \textbf{0.00}   & \textbf{0.00}   & \textbf{6.00}  & \textbf{1.50}   & \textbf{0.50}   & \textbf{0.00}   & \textbf{8.25}  & \textbf{10.50} & \textbf{1.00}   & \textbf{30.00}  \\ \bottomrule
\end{tabular}}
\caption{Accuracy and ASR of UltraClean against adaptive attacks (HTBD)}
\label{ASR_accuracy_HTBD_different_settings}
\end{table*}


\textbf{HTBD.} For HTBD attacks, there are two crucial parameters: the maximum perturbation $\epsilon$ and the trigger size ($\mathcal{T}_s$), which affect the performance and stealthiness of poisoned samples. Attackers may try to evade or break (i.e., achieve a higher ASR in the presence of detection) the defense of UltraClean by adjusting these parameters. Here we vary these two parameters to study their effects on UltraClean. Results are presented in Tables~\ref{HTBD_different_settings} and ~\ref{ASR_accuracy_HTBD_different_settings}. Recall that HTBD generates poisoned samples by adding perturbations to poisoned images to match source images' representations in the feature space. Therefore, $\epsilon$ in the HTBD is akin to the adversarial perturbations of the LCBD, which is the upper bound of perturbation imposed on the image. The trigger patched on source images affects their representations in the feature space. Thus the trigger size is also an important parameter for poisoned data generation. According to~\cite{DBLP:conf/aaai/SahaSP20}, a larger trigger size means larger perturbations added on poisoned images and always leads to better attack efficiency. As shown in the results, \textbf{UltraClean shows highly consistent performance against all adaptive attacks of different perturbation and trigger size.} In all experiments, UltraClean detects almost all poisoned samples and reduces the post-clean ASR to 0\% in most cases.

\section{Comparison to Other SOTA Defenses}
We have shown that UltraClean outperforms STRIP and SVD, which are poison-filtering-based defenses. To better demonstrate the effectiveness of UltraClean, we compare it to other SOTA defenses ABL~\cite{DBLP:journals/corr/abs-2110-11571} and ANP~\cite{wu2021adversarial}, which are \textbf{not} poison-filtering-based. Results are summarized in Table~\ref{tab:ABL}. ABL is a poison suppression-based defense that eliminates backdoors by unlearning the poisoned samples in the dataset. ANP is a model reconstruction-based defense that prunes some sensitive neurons to purify the injected backdoor. We conduct the experiment against BadNets on CIFAR-10. It can be seen that UltraClean achieves comparable clean accuracy and ASR while cleansing the dataset for reusability, as opposed to ABL and ANP that do not remove poisoned data.

\begin{table}[htbp]
\centering
\small

\begin{tabular}{c|ccc}
\hline
                    & \textbf{Acc.} & \textbf{ASR} & \textbf{BDR} \\ \hline
\textbf{Pre-Clean}  & 87.73\%       & 99.98\%            & -            \\
\textbf{ABL}~\cite{DBLP:journals/corr/abs-2110-11571}        & 89.03\%       & 0.00\%          & -            \\
\textbf{UltraClean} &  85.00\%             &    1.59\%          &   97.40\%           \\ \hline
\textbf{Pre-Clean}  &93.51 \%       & 99.92\%            & -            \\
\textbf{ANP}~\cite{wu2021adversarial}        & 90.20\%       & 0.45\%          & -            \\
\textbf{UltraClean} &  93.61\%             &     0.00\%         &   100\%           \\ \hline
\end{tabular}
\caption{Comparison of UltraClean to ABL against LCBD attack and ANP against BadNets on CIFAR-10 using ResNet.}
\label{tab:ABL}
\end{table}

\section{Effectiveness against Sleeper Agent}
We evaluate the effectiveness of UltraClean against various prestigious attacks. We notice the recently proposed clean-label backdoor attack Sleeper Agent (SA)~\cite{souri2022sleeper} employs a different attack mechanism from previous attacks, which crafts poisoned samples via gradient matching~\cite{DBLP:conf/iclr/GeipingFHCT0G21}. We evaluate UltraClean's performance against this latest type of attack in Table~\ref{tab:Sleeper}. We carry out the experiment on CIFAR-10 using ResNet-18. It can be seen that UltraClean still significantly reduces the ASR and successfully mitigate the backdoor effect, which again validates that UltraClean is effective against both clean and dirty-label attacks regardless of attack mechanisms. 

\begin{table}[htbp]
\centering
\small
\begin{tabular}{c|ccc}
\hline
 \textbf{}                      & \textbf{Acc.} & \textbf{ASR} & \textbf{BDR} \\ \hline
\textbf{Sleeper Agent}~\cite{souri2022sleeper} & 92.31\%       & 85.27\%      & -            \\
\textbf{UltraClean}    &   91.50\%            &   5.66\%           &   50.60\%           \\ \hline
\end{tabular}
\caption{Performance of UltraClean against Sleeper Agent attack on CIFAR-10 using ResNet-18.}
\label{tab:Sleeper}
\end{table}

\section{Study of Other Deonising Methods}
\label{sec: ablation_denoising}
We evaluate other denoising methods and present the results in Table~\ref{tab:denoising}. We perform the experiment against BadNets on CIFAR-10 using VGG16. Compared to our current method that employs local median and non-local mean denoising, incorporating more denoising methods only slightly improves or even degrades the performance. Therefore, we argue that it is unnecessary to introduce more denoising methods that may significantly increase the complexity of our defense.

\begin{table}[htbp]
\centering
\small
\begin{tabular}{c|ccc}
\hline
             & \textbf{Ours} & \textbf{Ours+Gaussian} & \textbf{Ours+Bilateral} \\ \hline
\textbf{BDR} & 94.42\%       & 95.16\%                & 83.80\%                 \\
\textbf{ASR} & 0.89\%        & 0.76\%                 & 100\%                   \\ \hline
\end{tabular}
\caption{Comparison of denoising methods against BadNets on CIFAR10 using VGG16.}
\label{tab:denoising}
\end{table}

\begin{table}[htbp]
\centering
\small
\setlength{\tabcolsep}{8pt} 
\resizebox{\linewidth}{!}{

\begin{tabular}{c|cccc}
\toprule
\textbf{}                                                                 & 
\textbf{\begin{tabular}[c]{@{}c@{}}GTSRB  \end{tabular}} & \textbf{\begin{tabular}[c]{@{}c@{}}CIFAR-10\end{tabular}} & \textbf{\begin{tabular}[c]{@{}c@{}}ImageNet\\ \end{tabular}} \\ \hline
\textbf{\begin{tabular}[c]{@{}c@{}}Total Time \end{tabular}}    & $\sim$398 (s)                                                  & $\sim$262 (s)                                                      & $\sim$57 (hrs)                                                    \\\hline
\textbf{\begin{tabular}[c]{@{}c@{}}\#Training Images\end{tabular}}      & 4,772                                                           & 50,000                                                              & 1,081,167                                                            \\\bottomrule
\end{tabular}}
\caption{Time consumption of detection over different datasets}
\label{time_cost}
\end{table}

\section{Time Cost of Detection}
We also analyze the efficiency of UltraClean and present the time consumption over each attack and dataset on our machine in Table~\ref{time_cost}. It can be seen that the proposed denoising operations are fast to process images on different datasets. Even for the ImageNet dataset, it only takes less than 0.2 seconds to process one image and about 57 hours to process the entire dataset (there are way more images than other datasets). Note that all the data are processed with each denoising operation in sequence; applying parallel processing may significantly reduce the time cost.

\end{document}